\documentclass[traditabstract]{aa_arxiv} 

\usepackage{graphicx}
\usepackage{txfonts}
\begin{document}
   \title{Transformation of the angular power spectrum of the Cosmic
     Microwave Background (CMB) radiation into reciprocal spaces and
     consequences of this approach} 

   \subtitle{ }

   \author{L. \v{C}ervinka}

   \institute{Institute of Physics AS CR, 
              Cukrovarnick\'{a}~10,~162~53~Praha 6, Czech Republic\\
              \email{L.Cervinka@icaris.cz} }

   \date{\today}

\abstract {A formalism of solid state physics has been applied to
  provide an additional tool for the research of cosmological
  problems. It is demonstrated how this new approach could be useful
  in the analysis of the Cosmic Microwave Background (CMB) data. After
  a transformation of the anisotropy spectrum of relict radiation into
  a special two-fold reciprocal space it was possible to propose a
  simple and general description of the interaction of relict photons
  with the matter by a ``relict radiation factor''. This factor
  enabled us to process the transformed CMB anisotropy spectrum by a
  Fourier transform and thus arrive to a radial electron density
  distribution function (RDF) in a reciprocal space. As a consequence
  it was possible to estimate distances between Objects of the order
  $\sim$10$^{2}$ [m] and the density of the ordinary matter
  $\sim$10$^{-22}$ [kg.m$^{-3}$]. Another analysis based on a direct
  calculation of the CMB radiation spectrum after its transformation
  into a simple reciprocal space and combined with appropriate
  structure modelling confirmed the cluster structure. The internal
  structure of Objects may be formed by Clusters distant 12
  [cm], whereas the internal structure of a Cluster consisted of
  particles distant $\sim$0.3 [nm].  This work points unequivocally to
  clustering processes and to a cluster-like structure of the matter
  and thus contributes to the understanding of the structure of
  density fluctuations. Simultaneously it sheds more light on the
  structure of the universe in the moment when the universe became
  transparent for photons. Clustering may be at the same time a new
  physical effect which has not been taken fully into consideration in
  the past. On the basis of our quantitative considerations it was
  possible to estimate the number of particles (protons, helium
  nuclei, electrons and other particles) in Objects and Clusters and
  the number of Clusters in an Object.}


\titlerunning{Transformation of CMB radiation into reciprocal spaces}

\maketitle


\section{Introduction}

The angular power spectrum (anisotropy spectrum) of the Cosmic
Microwave Background (CMB) radiation (Sievers 2003; Hinshaw 2003)
shows incredible similarity with X-ray or neutron scattering measured
on non-crystalline materials (\v{C}ervinka 1998, \v{C}ervinka et al. 2005),
see Figs. 1 and 2.  Astronomers ascribe to various peaks of the
anisotropy spectrum of the CMB radiation different processes (Hu
1995): It is the Sachs-Wolf effect, Doppler effect, Silk damping,
Rees-Sciama effect, Sunyaev-Zeldovich effect, etc. In this connection
it should be stated that all theoretical predictions of the standard
cosmological model are in very good agreement with the course of the
anisotropy spectrum of CMB radiation.  However, the formal similarity
in the form of both figures initiates the tempting idea if an analysis
of the anisotropy spectrum of relict radiation using an analogous
approach as is common in solid state physics, i.e. in the structural
analysis of disordered materials, would bring more information on the
structure of the early universe.

   \begin{figure}
   \centering
   \includegraphics[width=8cm]{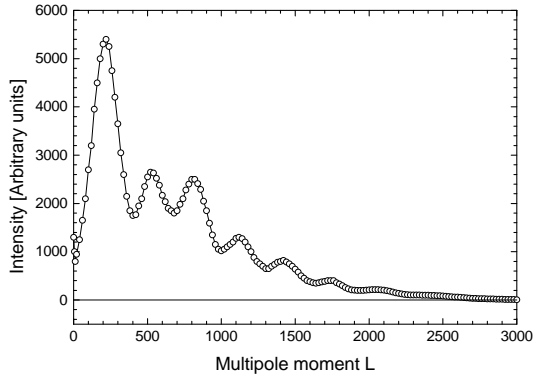}
      \caption{Anisotropy spectrum of the CMB radiation (Sievers
        2003). The figure describes the dependence of the magnitude of
        the intensity of microwave background on the multipole moment
        $L = 180^{\circ}/\alpha$, where $\alpha$ is the angle between
        two points in which the temperature fluctuations are compared
        to an overall medium temperature. The description of the
        Y-axis is for simplicity described in [Arbitrary units]. The
        original description was given as $L(L+1)C_{L}/2\pi$ in
        [$\mu$K$^{2}$] units, where $L$ is the multipole moment,
        $C_{L}$ is a function reflecting the width of the window
        measuring the temperature fluctuations.  }
         \label{fig1}
   \end{figure}

   \begin{figure}
   \centering
\includegraphics[width=8cm]{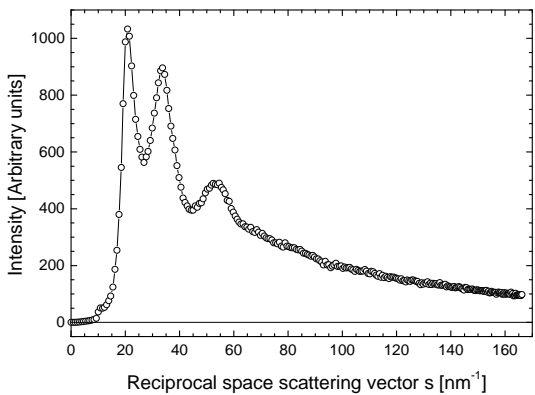}
      \caption{X-ray scattering diagram taken on a sample of a
        chalcogenide glass of a composition
        (Ge$_{0.19}$Ag$_{0.25}$Se$_{0.50}$) using the
        MoK$_{\alpha}$\ radiation, see \v{C}ervinka et al. (2005) for
        detail. The reciprocal space scattering vector $s$ is defined
        in equation (A.5).  }
         \label{fig2}
   \end{figure}

The inspiration for this approach we found further in the nowadays
situation: Although the individual disciplines in physics are highly
specialized, nevertheless their methods and results are shared in
areas that at the first sight may seem to be far apart. An example of
this is the already established use of elementary particle physics in
cosmology.

Similarly, we hope that it may be time now to apply the formalism of
solid state physics to some special cosmological problems and in this
way to provide an additional tool for their research. First of all our
new approach may be useful in the analysis of the CMB data. We will
show how after a transformation of the anisotropy spectrum of relict
radiation into a special two-fold reciprocal space we will be able to
process the transformed CMB anisotropy spectrum by a Fourier transform
and thus calculate a radial distribution function (RDF) of Objects in
a reciprocal space. Because the CMB radiation reflects the
fluctuations in the density of the matter, we hope that in this way
our study will be able to contribute to the understanding of the
structure of these density fluctuations (Sect.\ 3).

Moreover this work points quite unequivocally to clustering processes
and to a cluster-like structure of the matter, hence it sheds more
light on the structure of the universe in the moment when the universe
became transparent for photons (Sect.\ 4).

Clustering may be at the same time a new physical effect which has not
been taken fully into consideration in the past. On the basis of our
quantitative considerations it will be possible to derive the number
of particles (protons, helium nuclei, electrons) in Objects and
Clusters and of Clusters in an Object. This point will be demonstrated
in Sect.\ 4.2. and discussed in Sect.\ 5.1.

Another analysis based on a calculation of the CMB radiation spectrum
after its transformation into a simple reciprocal space, combined with
appropriate modelling experiments, will confirm the cluster structure
and indicate the differences between Objects and Clusters, see
Sects.\ 4.1. and 4.2.

Moreover, we will propose on the basis of this new formalism a general
description of the interaction of relict radiation with the matter. In
contrast to the atomic (coherent) and Compton (incoherent) scattering
factors calculated theoretically for all kinds of atoms in solid state
physics, in this special case we have generated a “relict radiation
factor” unifying all possible processes realized during the
interaction of relict radiation with various kinds of particles, see
Sects.\ 2.2.4. and 5.2.


\section{Construction of the Classic and Relict reciprocal space}

In solid state physics the principal mathematical method during the
structure analysis of the matter is the Fourier transform of the
intensity e.g. of X-rays or neutrons scattered by atoms building the
material. The experimental data are collected in the reciprocal space
and their Fourier transform brings the required information on the
distribution of atoms in the real space. Now we will try to apply this
approach to the CMB spectrum (see Fig. 1) and simultaneously point out
the complications we have to overcome in this direction.

The necessary basic mathematical apparatus is summarized in the
Appendix, the most important basic equations for the analysis of
“scattered” radiation and leading to the radial density distribution
function (RDF) are equations (A.1) and (A.2). The essential difference
in the use of terms ``scattering'' and ``interaction'' of photons will
be elucidated in the next Sect.\ 2.1.1.


\subsection{Discussion of parameters necessary for the calculation of
  a radial electron density distribution}


\subsubsection{The relict radiation factor}

During a conventional structure analysis with X-rays or neutrons, the
X-ray or neutron atomic scattering factors are a precise picture of
the interaction of radiation with the matter and are known precisely
(Wilson \& Price 1999). They enter into the calculation of the RDF in
correspondence with the composition of the studied material; see
equations (A.6), (A.7) and (A.10). Generally, for coherent scattering,
the atomic scattering factor $f$ is the ratio of the amplitude of
X-rays scattered by a given atom $E_{a}$ and that scattered according
to the classical theory by one single electron $E_{e}$, i.e. $f =
E_{a} / E_{e}$\ ($f\le Z$), where $Z$ is the number of electrons in
the atom.

Moreover, there are scattering factors not only for the coherent but
also for the incoherent (Compton) type of scattering, see e.g. later
on Fig. 7.

In our study, however, the basic obstacle is that with CMB photons we
have not a classic scattering process of photons on atoms; i.e. a
process described in equations of the Appendix. There are not atoms,
there are particles only (e.g. baryons, electrons, etc.), which
participate in the formation of the structure of density
fluctuations. Therefore we will speak throughout this article about an
``interaction'' instead of ``scattering'' in all cases when instead of
the classic ``atomic scattering factor'' the new \emph{``relict
  radiation factor''} will be used.

It is true that a part of the interaction of photons with electrons
before the recombination may be realised as Thomson scattering
(elastic scattering of electromagnetic radiation by a free charged
particle, as described by classical electromagnetism)
\footnote{It is just the low-energy limit of Compton scattering: 
the particle kinetic energy and photon frequency are the same before 
and after the scattering, however this limit is valid as long as the 
photon energy is much less than the mass energy of the particle.}, 
but the physical background as well as the complex picture of physical
processes describing the interaction of relict photons with the non
uniform matter composed of various particles (electrons, ions, etc.)
is not known precisely.

It is therefore evident that it will not be possible to use the
conventional atomic scattering factors and that a \emph{new special
  factor} reflecting the complexity of interaction processes of
photons with the primordial matter has to be constructed. We only
point out that the description of these interactions is possible only
in a special two-fold reciprocal space into which the CMB spectrum is
transformed. This new factor will be called the \emph{relict radiation
  factor} and substitutes all complicated processes which participate
in the formation of the angular power spectrum of CMB radiation.

In this way this approach and-or formalism may also shed some new
light on the (well-known) physical processes taking place in the
primordial plasma.

In order to construct the relict radiation factor we will use a basic
mathematical criterion which serves well also in the classic case (see
also later on Sect.\ 2.2.4.): Only when the atomic coherent and
incoherent scattering factors (for X-rays or neutrons) are included
into the calculation correctly, then the Fourier transform of the
quantity $i(s)$ according equation (A.2) presents data without or with
minimal parasitic fluctuations. Similarly also in this case this
criterion should help us during the construction of the relict
radiation factor: The relict radiation factor has to be constructed in
such a way that after its insertion into the calculation of the RDF
(see equations (A.1) and (A.6), where the relict radiation factor is
then labelled $f_{m}$, possible parasitic fluctuations on the RDF
$\rho(r)$ should be again minimized to the greatest possible extent.

The construction of the relict radiation factor is presented in
Sect.\ 2.2.4.


\subsubsection{The wavelength of radiation}

The wavelength of radiation is a quantity of highest importance,
too. It follows from equation (A.5), that the greater the wavelength
the smaller is the maximal possible value $s_{\mathrm{max}}$\ of the
reciprocal space vector. At the same time the upper limit of the
integral in equation (A.2) strongly influences the quality of the
Fourier transform.

Although there is a broad distribution of wavelengths of photons (see
later on the discussion in Sect.\ 5.3.) the calculation will be
undertaken for the wavelength corresponding to the maximum of the
wavelength distribution, i.e. for the wavelength $\lambda$ = 1.9 [mm]
corresponding to the temperature 2.725 [K] of the Universe today.

That this wavelength is rational is based on three arguments. First of
all photons with this wavelength bring us \emph{today} the information
on their last several interactions with particles, in the second place
the CMB radiation spectrum is the same for all wavelengths and in the
third place the wavelength corresponding to the maximum of the
wavelength distribution secures the highest probability of the
interaction process of photons with the matter.


\subsubsection{The macroscopic density}

The macroscopic density is a parameter which contributes to the
calculation of the first expression on the right side of equation
(A.2), i.e. it characterises ``the slope'' of the total disorder, see
e.g. Fig.8. The calculation only of the second member of equation
(A.2) may help in an estimate of this quantity, because it is highly
improbable that oscillations on a properly calculated RDF should be
negative. This fact has been used when estimating the density of the
matter, see Sect.\ 3.2. It is important that the basic features of a
RDF (positions of coordination spheres) are already determined by the
calculation only of the second member in equation (A.2).


\subsection{Preparatory calculations}

\subsubsection{The Classic reciprocal space}

During a classic scattering experiment we measure the intensity of the
scattered radiation (e.g. X-rays) as a function of the scattering
angle $\theta_{\mathrm{Classic}}$. This scattering angle describes in
real space the angle between the incident and scattered radiation. Its
relation with the scattering vector in reciprocal space was described
in equation (A.5).

On the other hand the angle $\alpha$\ in the anisotropy spectrum of
relict radiation (see already Fig. 1) is not a scattering angle. It is
an angle characterizing a distance between an arbitrary point to
another - in those different points the temperature fluctuation is
measured and compared with the overall medium one.

In order to overcome the incomparableness between the angles
$\theta$\ and $\alpha$\ we will construct \emph{an angle dependent
  reciprocal space to the angle $\alpha$}. The basic quantity
determining this reciprocal space will be the scattering angle
$\theta_{\mathrm{Classic}}$.

We will suppose that the maximum possible value of the classic
scattering angle $\theta^{\mathrm{max}}_{\mathrm{Classic}} =
90^{\circ}$\, corresponds to the maximum value of the multipole moment
$L^{\mathrm{max}}$=3000.

As a consequence we receive a transformation coefficient $Q$
\begin{equation}
\theta^{\mathrm{max}}_{\mathrm{Classic}} = Q \, L^{\mathrm{max}}
\label{eq1}
\end{equation}
(its value in this case is $Q$ = 0.03).  We are then able to calculate
the whole set of $\theta_{\mathrm{Classic}}$ angles
\begin{equation}
\theta_{\mathrm{Classic}} \, = \, L \, Q \, = \, L \:
(\theta^{\mathrm{max}}_{\mathrm{Classic}}/L^{\mathrm{max}})
\label{eq2}
\end{equation}
and because $L$ = 180/$\alpha$, then
\begin{equation}
\theta_{\mathrm{Classic}} = \frac{1}{\alpha}  \, 180 \: 
(\theta^{\mathrm{max}}_{\mathrm{Classic}}/L^{\mathrm{max}}) \;\; ,
\label{eq3}
\end{equation}
i.e.
\begin{equation}
\theta_{\mathrm{Classic}} = \frac{1}{\alpha}  \, P_{\mathrm{Classic}}
\label{eq4}
\end{equation}
where
\begin{equation}
P_{\mathrm{Classic}} = 180\: Q\: [\mbox{deg}^{2}]
\label{eq5}
\end{equation}
is a coefficient enabling the transition between space $\alpha$ and
the space $\theta_{\mathrm{Classic}}$ and where \emph{the angular
  space $\theta_{\mathrm{Classic}}$ is reciprocal to the angular space
  $\alpha$ .}

According equation (A.5) we are now able to construct the whole set of
scattering vectors $s_{\mathrm{Classic}}$
\begin{equation}
s_{\mathrm{Classic}} = 4 \pi \, \frac{\sin
  \theta_{\mathrm{Classic}}}{\lambda} \;\; ,
\label{eq6}
\end{equation}
where $\lambda$ is the wavelength of the relict radiation. It should
be noted that the quantities $s_{\mathrm{Classic}}$ and $\alpha$ are
in an indirect relation. The space of the vector
$s_{\mathrm{Classic}}$ will be further on called a \emph{ ``classic
  reciprocal space''}.

It should be pointed out that in this construction (see equation (6))
the scattering vector $s_{\mathrm{Classic}}$ is defined in the
reciprocal space (1/$\lambda$) and that \emph{this space is now dipped
  into the reciprocal space (1/$\alpha$)}, see equations (2), (4) and
(6). For this ``dipping'' we will use further on the expression that
\emph{the space $s_{\mathrm{Classic}}$\ is a 2-fold reciprocal space
  to the space} $\alpha$.

The recalculation of the original data presented in Fig. 1 using
equations (4) and (6) is shown in Fig. 3. This new intensity
dependence is labelled $I_{\mathrm{Classic}}(s_{\mathrm{Classic}})$.

   \begin{figure}
   \centering
\includegraphics[width=8cm]{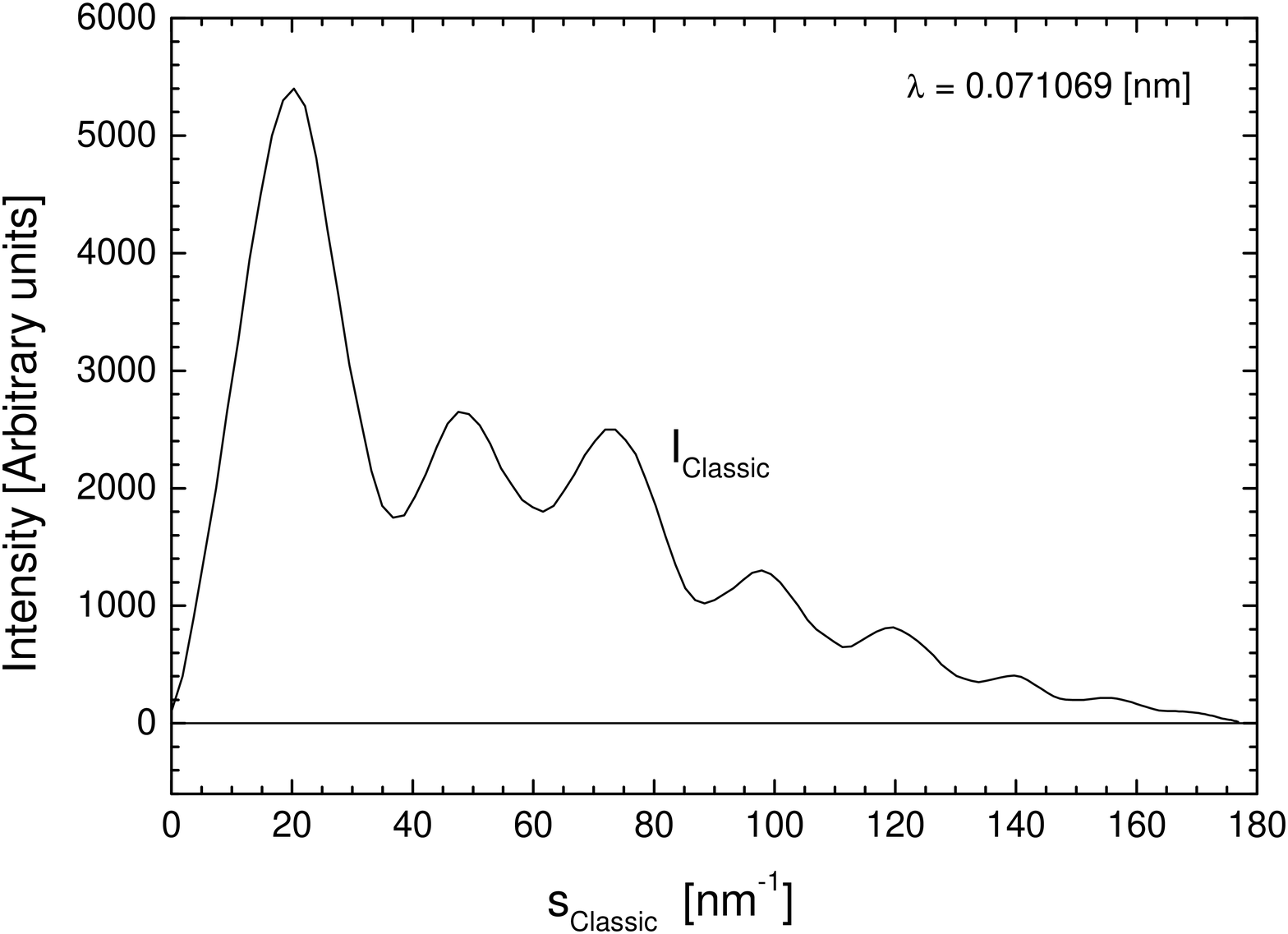}
      \caption{Anisotropy spectrum of the relict radiation shown in
        Fig. 1 is recalculated as a function of
        $s_{\mathrm{Classic}}$, i.e. after a rescaling of the angular
        moment $L$ into the dependence
        $I_{\mathrm{Classic}}(s_{\mathrm{Classic}})$.  The rescaling
        of the angular moment $L$ is
        realized on the basis of equations (2), (4) and (6) and using
        the MoK$_{\alpha}$\ radiation wavelength $\lambda$=0.071609
        [nm].  }
         \label{fig3}
   \end{figure}


\subsubsection{The Relict reciprocal space}

There is a possibility to construct another reciprocal space which
will be based directly on the angle $\alpha$. For a better comparison
and lucidity we will use now for the angle $\alpha$ the labelling
$\theta_{\mathrm{Relict}}$\ i.e.
\begin{equation}
\alpha = \theta_{\mathrm{Relict}} \; ,
\label{eq7}
\end{equation}
hence 
\begin{equation}
L = 180^{\circ}/\alpha = 180^{\circ}/\theta_{\mathrm{Relict}}.
\label{eq8}
\end{equation}

In close analogy with equation (A.5) we now transform the anisotropy
spectrum of CMB (relict) radiation into a reciprocal space
(1/$\lambda$) described by the parameter $S_{\mathrm{Relict}}$
\begin{equation}
S_{\mathrm{Relict}} = 4 \pi \, \frac{\sin \theta_{\mathrm{Relict}}}{\lambda}
\label{eq9}
\end{equation}
where $\lambda$ is the wavelength of the relict radiation. The space
of the vector $S_{\mathrm{Relict}}$ will be further on called the
\emph{``Relict reciprocal space''}.

It should be noted that \emph{quantities $S_{\mathrm{Relict}}$ and
  $\theta_{\mathrm{Relict}}$ = $\alpha$ are in a direct relation}. The
anisotropy spectrum of the CMB radiation rescaled on the basis of
equation (9) is here labelled
$I_{\mathrm{Relict}}(S_{\mathrm{Relict}})$ and is shown in Fig. 4.

   \begin{figure}
   \centering
   \includegraphics[width=8cm]{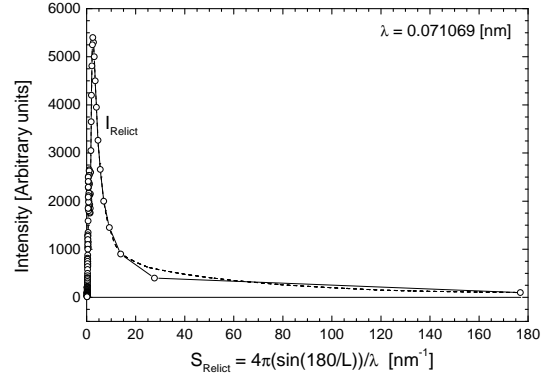}
      \caption{Anisotropy spectrum of the relict radiation shown in
        Fig. 1 is after a rescaling of the angular moment $L$,
        recalculated as a function of the Relict reciprocal space
        vector $S_{\mathrm{Relict}}$, into the dependence
        $I_{\mathrm{Relict}}(S_{\mathrm{Relict}})$. The rescaling of
        the angular moment $L$ is realized on the basis of equations
        (8), (9) and (11) using the MoK$_{\alpha}$\ radiation
        wavelength $\lambda$\ = 0.071609 [nm]. The dashed line
        represents a smoothed curve.  }
         \label{fig4}
   \end{figure}


\subsubsection{Relation between the Classic and Relict reciprocal
  space} 

The Classic reciprocal space was defined in equation (6), which can be
rewritten, using equation (2) into an $L$-dependent form
\begin{equation}
s_{\mathrm{Classic}} = 4 \pi \, \frac{\sin (L Q )}{\lambda}
\label{eq10}
\end{equation}

Similarly the Relict reciprocal space was defined in equation (9),
which can be rewritten using equation (8) also into an $L$-dependent
form
\begin{equation}
S_{\mathrm{Relict}}
= 4 \pi \, \frac{\sin(180^{\circ}/L)}{\lambda}
\label{eq11}
\end{equation}

In Fig. 5 we show the dependencies $s_{\mathrm{Classic}}(L)$\ and
$S_{\mathrm{Relict}}(L)$ simultaneously with the function
1/$S_{\mathrm{Relict}}(L)$.

   \begin{figure}
   \centering
   \includegraphics[width=8cm]{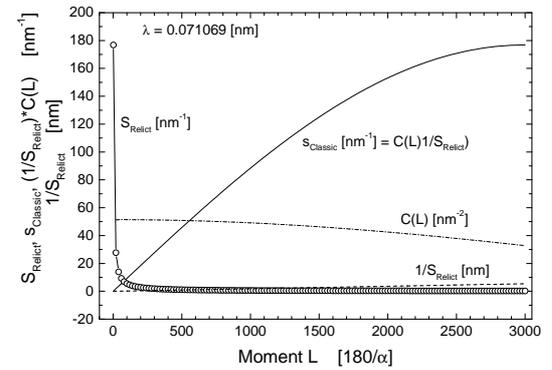}
      \caption{Mutual behaviour of reciprocal space vectors
        $S_{\mathrm{Relict}}$ (empty circles - equation (11)) and
        $s_{\mathrm{Classic}}$ (full line - equation (10)) is
        shown. The dependence of 1/$S_{\mathrm{Relict}}$\ (dashed
        line) on $L$ is linear. Multiplication of
        1/$S_{\mathrm{Relict}}$ by the coefficient $C(L)$, see
        equation (12), produces a curve lying precisely on the line
        $s_{\mathrm{Classic}}$. All calculations are done for the
        wavelength $\lambda$\ = 0.071069 [nm]. See text for detail.  }
         \label{fig5}
   \end{figure}

Now it is possible to find an $L$-dependent transformation coefficient
$C(L)$ for which
\begin{equation}
s_{\mathrm{Classic}} \, = \, C(L) \, \frac{1}{S_{\mathrm{Relict}}}
\;\; ,
\label{eq12}
\end{equation}
where e.g. for the wavelength $\lambda$\ = 0.071069 [nm] the
coefficient $C(L)$ (dimensionality [nm$^{-2}$]) has the course
visualised in Fig. 5. In reality the coefficient $C$ is not only a
function of $L$ but simultaneously a function of $\lambda$,
i.e. precisely it should be written as $C(L,\lambda)$.
\footnote{That $C$ is a function of $\lambda$ then indicates
that for every wavelength there has to be another calculation
of equation (12) and simultaneously there has to be another
calculation of equations (6) and (9).}

Equation (12) is important, because it enables the mutual comparison
of the Classic reciprocal space (represented by vector
$s_{\mathrm{Classic}}$) with the Relict reciprocal space (represented
by vector $S_{\mathrm{Relict}}$) and vice versa. To summarize:\emph{
  the mutual relationship between the Classic reciprocal space and the
  Relict reciprocal space is reciprocal.}


\subsubsection{Construction of the relict radiation factor}

Generally, a correct scattering factor has to fulfil three criterions:

(A) the $I_{\mathrm{norm}}(s)$ curve should oscillate along the
$I_{\mathrm{gas}}(s)$\ curve and as a consequence according equation
(A.9)

(B) the curve $I_{\mathrm{distr}}(s)$ should oscillate along the zero value of
the intensity axis;

(C) the resulting RDF must not be contaminated by parasitic
fluctuations due to bad scaling (see Sect.\ A2.) as a consequence of a
bad course of the scattering factor.

The mutual relation between quantities $I_{\mathrm{norm}}(s)$,
$I_{\mathrm{gas}}(s)$\ and $I_{\mathrm{distr}}(s)$\ is explained in the
Appendix, see equations (A.9), (A.10) and (B.1) with (B.2).

   \begin{figure}
   \centering
   \includegraphics[width=8cm]{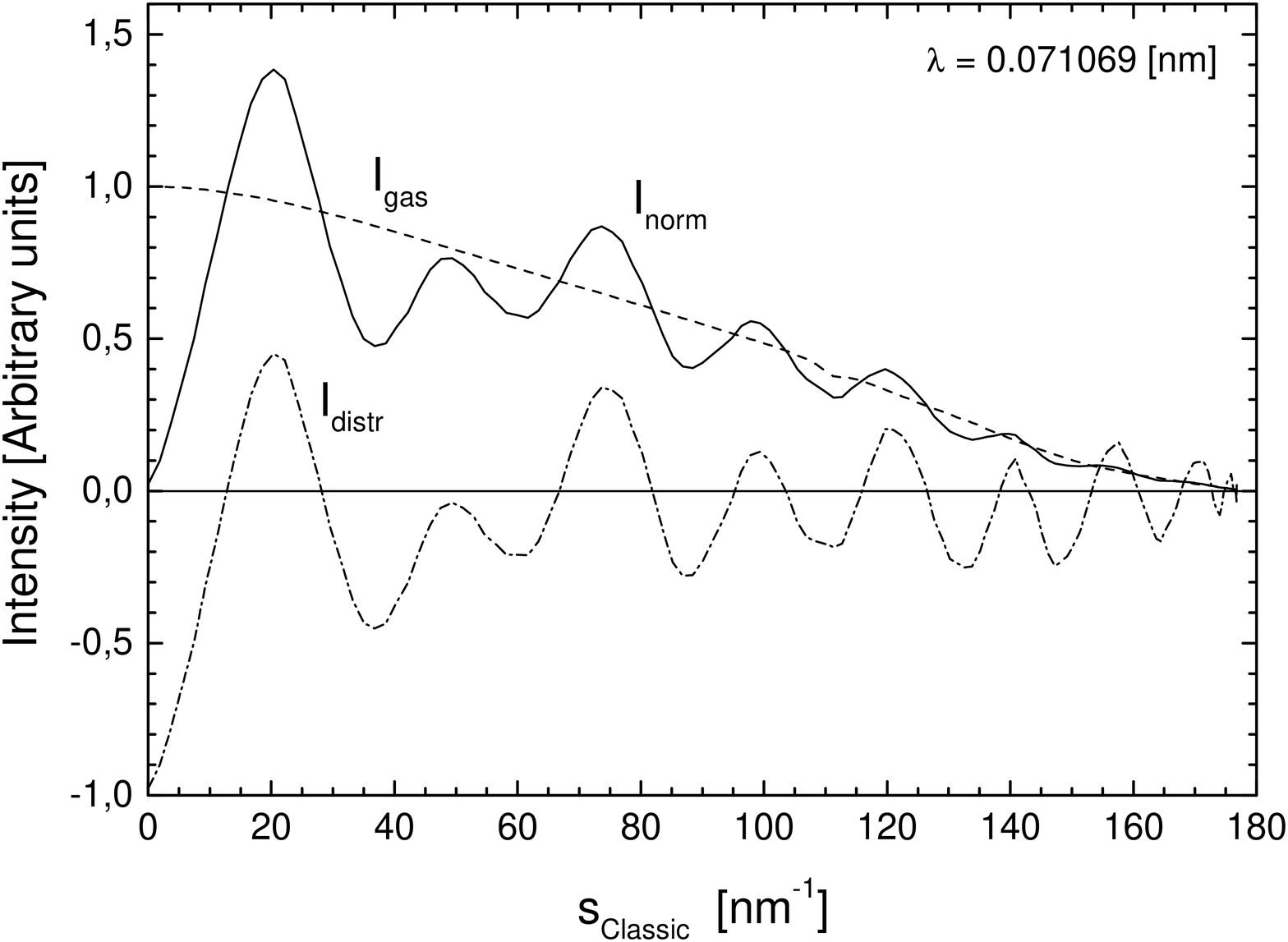}
      \caption{Calculation of quantities $I_{\mathrm{norm}}(s)$\ -
        full line, $I_{\mathrm{gas}}(s)$\ - dashed line (see equation
        (14)) and of $I_{\mathrm{distr}}(s)$\ - dashed dotted line, according
        equations (A.9),(A.10),(B.1) and (B.2) using the
        ``artificial'' relict radiation factor $f_{\mathrm{Relict}}$
        for the wavelength $\lambda$ = 0.071069 [nm]. Oscillations of
        the curve $I_{\mathrm{distr}}(s)$\ are along the x-axis; hence all the
        criterions set at the beginning of this section are
        fulfilled. See text for details.  }
         \label{fig6}
   \end{figure}

In Fig. 6 the calculation of the crucial curve $I_{\mathrm{gas}}$ is
undertaken for the relict radiation factor $f_{\mathrm{Relict}}$. The
form of this factor was determined by the trial and error method and
is shown in Fig. 7. In this figure is the factor $f_{\mathrm{Relict}}$
compared with the coherent ($f^{\mathrm{X}}_{\mathrm{coh}}$) and
incoherent ($f^{\mathrm{X}}_{\mathrm{incoh}}$) atomic scattering
factors for X-rays corresponding to the Hydrogen atom (Wilson \& Price
1999).

Similarly as for X-rays we have set the relict radiation factor
$f_{\mathrm{Relict}}$
\begin{equation}
 f_{\mathrm{Relict}}=1 \quad\mbox{for}\quad  s=0. 
\label{eq13}
\end{equation}
and further, we have set in equation (A.7) $Z$ = 1 and $m$ = 1, hence
in equation (A.6) is $K_{m}$ = 1. From this point of view our
construction of the relict radiation factor $f_{\mathrm{Relict}}$
should formally correspond to a \emph{``hydrogen-like'' particle}.

Further we have to point out that in connection with the presentation
of the quantity $I_{\mathrm{gas}}(s)$ in equation (A.10) its course in
Fig. 6 is given now by the relation
\begin{equation}
I_{\mathrm{gas}}(s) = \sum_{m} a_{m} f^{2}_{\mathrm{Relict}}  \;\; .
\label{eq14}
\end{equation}

In Fig. 6 we see that the function $I_{\mathrm{norm}}(s)$ is properly
oscillating along the function $I_{\mathrm{gas}}(s)$\ and therefore
the function $I_{\mathrm{distr}}(s)$ is properly oscillating along the zero
line.  The consequence is that we will obtain a ``proper'' radial
distribution function, i.e. without any parasitic maxima, see
Sect.\ 3.1.

   \begin{figure}
   \centering
   \includegraphics[width=8cm]{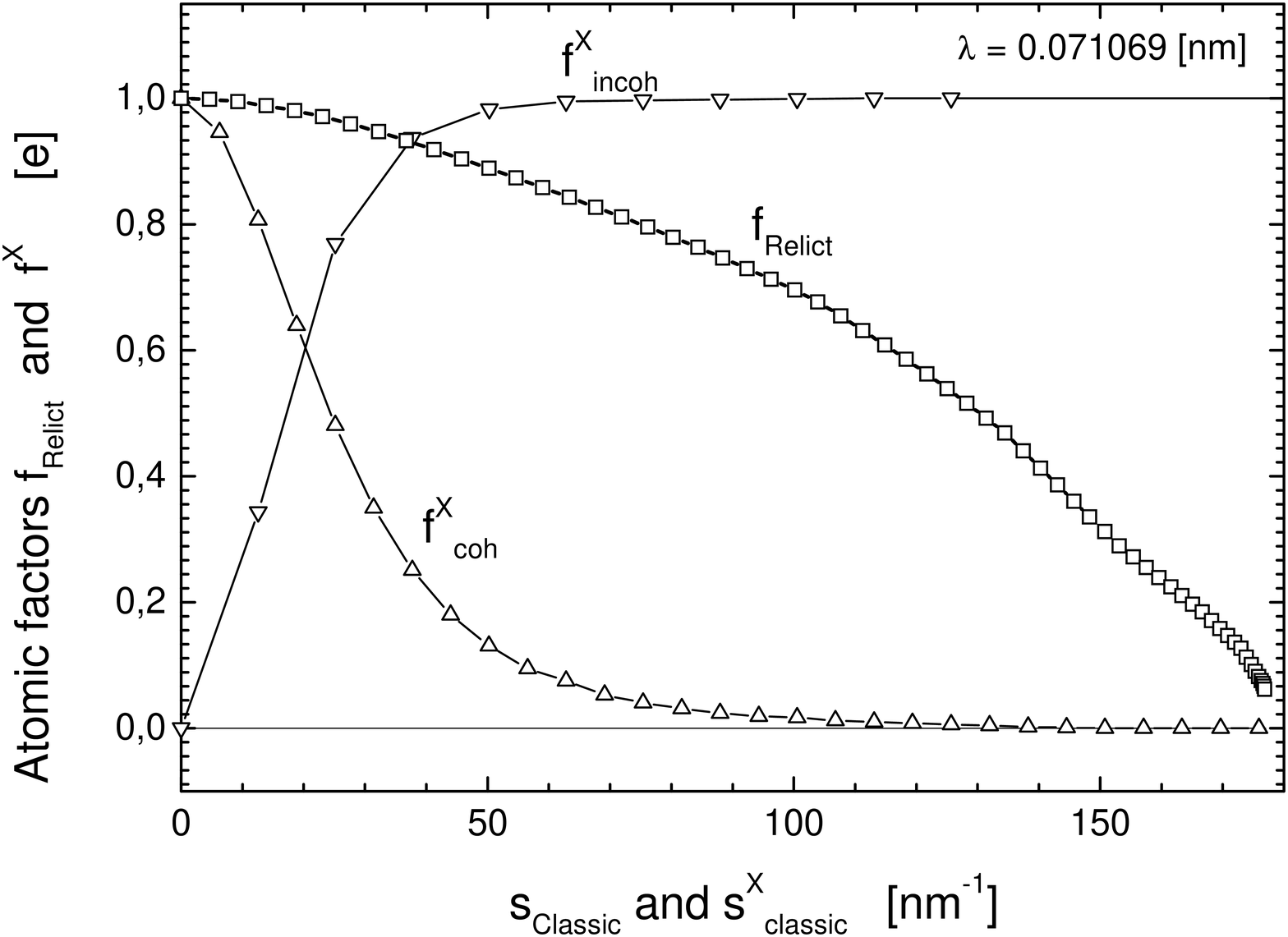}
      \caption{Behaviour of the relict radiation factor
        $f_{\mathrm{Relict}}$ is shown. For comparison the courses of
        the classic coherent and incoherent X-ray atomic scattering
        factors $f^{\mathrm{X}_{\mathrm{coh}}}$ and
        $f^{\mathrm{X}_{\mathrm{incoh}}}$ for Hydrogen are
        included. The parameter $s_{\mathrm{Classic}}$ is defined in
        equation (6), the parameter $s^{X}_{\mathrm{Classic}}$ is
        described in equation (A.5). Data for
        $f^{\mathrm{X}_{\mathrm{coh}}}$ and
        $f^{\mathrm{X}_{\mathrm{incoh}}}$ are taken from Wilson \&
        Price 1999. The calculation is demonstrated for the wavelength
        $\lambda$ = 0.071069 [nm].  }
         \label{fig7}
   \end{figure}


\subsubsection{Relation between the Classic and Relict distribution of
  distances} 

We rewrite now the basic equation (A.2) using the scattering vector in
the Classic reciprocal space $s_{\mathrm{Classic}}$, see equation (6)
\begin{equation}
\rho( r [\mathrm{nm}^{*}]) = \rho_{\mathrm{0}}^{\mathrm{Medium}}(r) \, + \, 
\rho^{\mathrm{Fourier}}(r,I(s_{\mathrm{Classic}})) \;\; ,
\label{eq15}
\end{equation}
where $\rho_{0}^{\mathrm{Medium}}(r)$ is the member which is not
Fourier-dependent and describes the structure-less total disorder
depending on the density of the matter.

The parameter $r$ is measured in [nm*] in order to emphasize that the
calculation of the RDF $\rho(r)$\ is realized on the basis of the
parameter $s_{\mathrm{Classic}}$, which is dipped in a 2-fold
reciprocal space (see Sect.\ 2.2.1.). In other words: the calculation
of the RDF $\rho(r)$\ is realized \emph{in the}
\emph{\textbf{reciprocal space}} \emph{ of classic distances}, which
have the dimension [nm*].  Here we again point out the fact, that
\emph{classic distances} are distances between Objects calculated on
the basis of the function
$I_{\mathrm{Classic}}(s_{\mathrm{Classic}})$, see Fig. 3, which we
analyze using equation (15).

In order to receive now the information in the \emph{\textbf{real
space}} \emph{of classic distances} (characterized by the parameter
$R$) we must calculate the reciprocal value of the parameter $r$ ,
hence the relation between $r$\ and $R$ is   
\begin{equation}
\frac{1}{r [\mathrm{nm}^{*}]} = R \, [\mathrm{nm}] \;\; .
\label{eq16}
\end{equation}

It would be now possible to rewrite quite formally equation (A.2)
using the scattering vector in the Relict reciprocal space
$S_{\mathrm{Relict}}$, see equation (9). Similarly as for equation
(15) we would receive   
\begin{equation}
\rho(R) = \rho_{0}^{\mathrm{Medium}}(R) \, + \,
\rho^{\mathrm{Fourier}}(R,I(S_{\mathrm{Relict}})) \;\; .
\label{eq17}
\end{equation}

Quite hypothetically the RDF $\rho(r)$\ would then bring us
information on the \emph{real space of relict distances}, which have the
dimension [nm]. Actually, however, a RDF will not be calculated in
this case, because the distribution $I(S_{\mathrm{Relict}})$, see
Fig. 4, is not convenient for a Fourier transform. The calculation of
\emph{relict distances} in the real space, i.e. of distances between
complex Objects (big clusters) will be done on the basis of a
theoretical calculation of the function $I(S_{\mathrm{Relict}})$ using
the Debye formula (18) calculated for appropriate models, see later on
Sect.\ 4.


\section{Calculations in the Classic reciprocal space}

\subsection{Calculation of RDFs} 

In our first example we calculate in Fig. 8 the RDF of Objects
corresponding to the Fourier transform of intensities
$\rho^{\mathrm{Fourier}}(r,I(s_{\mathrm{Classic}}))$ for the
wavelength $\lambda$ = 0.071069 [nm], see equation (15). The scaling
of intensities has been already demonstrated in Fig. 6 on the basis of
the relict radiation factor $f_{\mathrm{Relict}}$\ constructed in
Fig. 7.

The calculated RDF shows a form typical for RDFs obtained for
disordered materials. It turns out that in the region from 0.1 to 0.4
[nm*] most essential are the maxima $^{1}r$\ and $^{2}r$\ separated by
a minimum $^{min}r$ , which are followed by a structure-less
course. Such behaviour indicates the existence of ordering in the
matter. In other words, there is a distinctive separation of the
matter ending its ordering by the sphere at 0.312 [nm*] from the
residual structure-less ordering starting with a plain peak at 0.395
[nm*]. The small maximum $^{0}r$\ located at 0.172 [nm*] we consider
for the present as irrelevant.

In the same way we calculated RDFs for four more wavelengths,
i.e. 0.110674 [nm] ($\lambda_{\mathrm{SeK\alpha}}$), 0.154178 [nm]
($\lambda_{\mathrm{CuK\alpha}}$), 0.250466 [nm] ($\lambda_{\mathrm{VK\alpha}}$)
and 0.537334 [nm] ($\lambda_{\mathrm{SK\alpha}}$).  From these calculations it
follows that, as expected, the dependence of the magnitude of
corresponding coordination spheres on the wavelength $\lambda$ is
linear, see Fig. 9, moreover, all RDFs have the same appearance.

   \begin{figure}
   \centering
   \includegraphics[width=8cm]{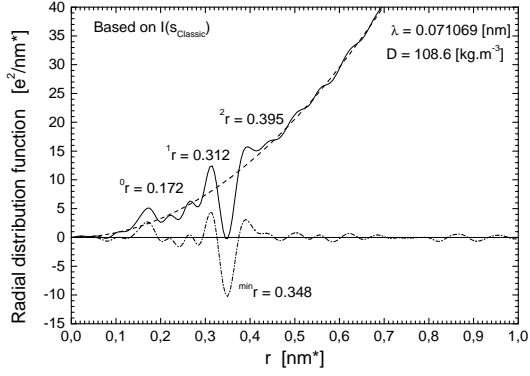}
      \caption{Calculation of the radial distribution function (RDF)
        according equation (15) for the wavelength $\lambda$\ =
        0.071069 [nm]. The dashed-dotted line corresponds to the
        second member in equation (15), the dashed line is the first
        member in this equation (dependent on density) and the full
        line is the sum of both components, see text for
        details. Value of the density $D$ necessary to shift the
        minimum at 0.348 [nm*] to positive values of the RDF is
        indicated in the upper right corner.  }
         \label{fig8}
   \end{figure}

   \begin{figure}
   \centering
   \includegraphics[width=8cm]{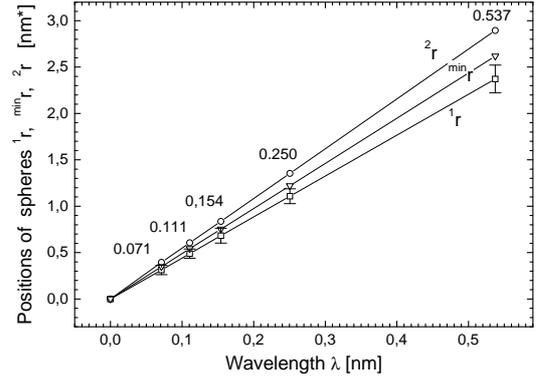}
      \caption{Dependences of most important Object distances, i.e. of
        coordination spheres $^{1}r$\ (squares), $^{2}r$\ (circles)
        separated by the minimum $^{\mathrm{min}}r$\ (down triangles) on the
        wavelength $\lambda$\ in the reciprocal space [nm*] according
        Fig. 8 and from analogical calculations for wavelengths
        0.110674, 0.154178, 0.250466 and 0.537334 [nm]. For an easier
        survey error bars are inserted only for the sphere $^{1}r$\ .
      }
         \label{fig9}
   \end{figure}

In this connection we have to point out, that the distances are
measured in reciprocal space distances [nm*] and that, with respect to
equation (16), these distances have to be recalculated to ``real space''
distances, e.g. in [km]. This recalculation is realized in Table 1,
where we review the results from all wavelengths (Fig. 9) and
simultaneously extrapolate the distances to the wavelength of relict
radiation photons $\lambda$=1.9 [mm].

Real space distances between Objects calculated in Table 1 are
visualized in Fig. 10. The extrapolation to the wavelength of relict
photons 1.9 [mm] indicates that for this wavelength the shortest
Object distances are in the range between 100 to 120 meters. In all
further considerations, however, we will use as the most distinctive
number characterizing the distance between Objects the value $^{2}R$
describing the start of the structure-less region, i.e. the distance
$^{2}R$ = 98$\pm$2 [m].

\begin {table*}
\caption{Review of most important nearest neighbour distances between
  Objects ($^{1}r$\ and $^{2}r$) separated by the minimum
  $^{\mathrm{min}}r$\ on the wavelength $\lambda$ (see
  Fig.~\ref{fig8}). Recalculation to real space distances $R$ is
  included. Simultaneously an extrapolation of distances corresponding
  to the wavelength of relict radiation photons 1.9 [mm] is computed
  together with an estimate of final errors.
\label{table:1}}
\centering
\begin{tabular}{ccccccc}
\hline\hline
 \multicolumn{4}{c}{Review
of reciprocal space distances in [nm*] on} &
 \multicolumn{3}{c}{Recalculation 
of reciprocal space distances [nm*]} \\
 \multicolumn{4}{c}{the basis of results
presented in Figs.~\ref{fig7}, \ref{fig8} and \ref{fig9}} &
 \multicolumn{3}{c}{between Objects into the real
space distances [km]} \\
$\lambda$ [nm] & $^{1}r$ [nm*] & $^{\mathrm{min}}r$ [nm*] & $^{2}r$ [nm*]
& $^{1}R$ [km] & $^{\mathrm{min}}R$ [km] & $^{2}R$ [km] \\
  &  &  &  & = 1/$^{1}r$ [nm$^{*-1}$] &
= 1/$^{\mathrm{min}}r$ [nm$^{*-1}$] & = 1/$^{2}r$ [nm$^{*-1}$] \\[0.5ex] 
\hline 
0.071069 & 0.312 & 0.348 & 0.395 & 3 205 128 & 2 873 563 & 2 538 071 \\
0.110674 & 0.488 & 0.542 & 0.605 & 2 049 180 & 1 845 018 & 1 652 893 \\
0.154178 & 0.682 & 0.752 & 0.836 & 1 466 276 & 1 329 787 & 1 136 172 \\
0.250466 & 1.107 & 1.221 & 1.353 & 903 342 & 819 001 & 739 098 \\
0.537334 & 2.372 & 2.618 & 2.895 & 421 585 & 381 971 & 345 423 \\[0.5ex]
\cline{1-4} 
\multicolumn{4}{c}{Extrapolation to higher wavelengths $\lambda$}
& \multicolumn{3}{c}{  } \\
\cline{1-4} 
1 & 4.42 & 4.87 & 5.39 & 226 072 & 205 231 & 185 694 \\
10 & 44 & 49 & 54 & 22 607 & 20 523 & 18 569 \\
100 & 442 & 487 & 539 & 2 261 & 2 052 & 1 857 \\
500 & 2 212 & 2 436 & 2 693 & 452 & 410 & 371 \\
1 000 & 4 423 & 4 873 & 5 385 & 226 & 205 & 186 \\
1 000 000 & 4 423 376 & 4 872 561 & 5 385 196 & 0.226 & 0.205 & 0.185 \\
1 900 000 & 8 404 414 & 9 257 865 & 10 231 872 & 0.119 & 0.108 & 0.098 \\[0.5ex]
\hline 
= 1.9 & = 8.4$\pm$0.1 & = 9.3$\pm$0.1 & = 10.2$\pm$0.1 &
= 119$\pm$2 & = 108$\pm$2 & = 98$\pm$2 \\
  \mbox{[mm]} & [mm$^{*}$] & [mm$^{*}$] & [mm$^{*}$] & [m] & [m] & [m] \\
\hline\hline
\end{tabular}
\end{table*}

   \begin{figure}
   \centering
   \includegraphics[width=8cm]{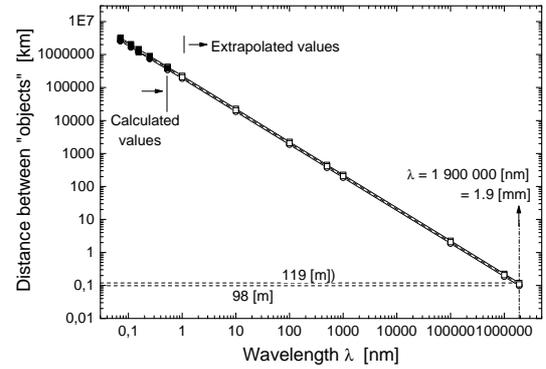}
      \caption{Dependence of real space distances between Objects,
        i.e. $^{1}R$ and $^{2}R$ (full squares and circles) separated
        by the minimum $^{\mathrm{min}}R$ (full up triangles) on the wavelength
        $\lambda$ (see Table 1 for details). Simultaneously an
        extrapolation to distances corresponding to the wavelength of
        CMB (relict) photons $\lambda$ = 1.9 [mm] is visualized (empty
        squares, circles and up triangles, respectively).  }
         \label{fig10}
   \end{figure}


\subsection{Calculation of the density}

The calculation presented in Fig. 8 and repeated for four additional
wavelengths enabled us to estimate the density of the matter, i.e. the
important parameter effecting the first member
$\rho^{\mathrm{Medium}}_{0}(r)$ in equation (15). We simply supposed
that the fluctuations of the RDF should not be negative. In order to
shift in Fig. 8 the minimum at $^{\mathrm{min}}r$\ = 0.348 [nm*] to positive
values we had to set the density to a value $D$ = 108.60
[kg.m$^{-3}$].  In the same way we have determined densities for the
remaining four wavelengths.

The results are summarized in Fig. 11 and Table 2. In the log-scale is
the dependence of density on the wavelength nearly linear and
therefore enables again an extrapolation to higher wavelengths. This
extrapolation is presented in Table 2 and visualized in Fig. 12.

It follows from Table 2 and Fig. 12 that the most probable medium
density of density fluctuations of the matter with which CMB (relict)
photons realized their last interaction is
$D$=9$\times$10$^{-23}$ [kg.m$^{-3}$]. Taking in account the
limits of our calculation then the density can be formally written as
$D\sim10^{-22}\pm10^{-3}$ [kg.m$^{-3}$].  see also Fig. 12 and Table
2.

   \begin{figure}
   \centering
   \includegraphics[width=8cm]{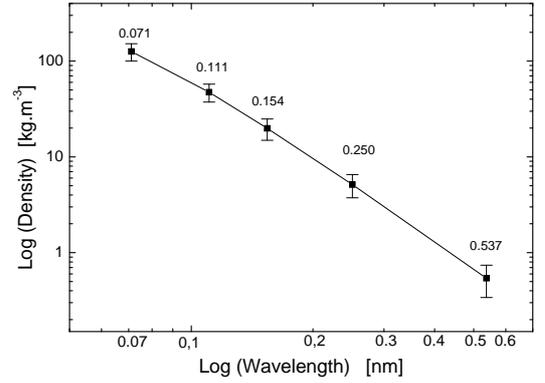}
      \caption{Dependence of macroscopic densities on short
        wavelengths. In the log-scale this dependence is nearly
        linear. Numerical values are given in Table 2. Numbers
        indicate wavelengths, for which the corresponding RDF was
        calculated, }
         \label{fig11}
   \end{figure}

\begin{table}
\caption{Review of numerical values of densities according
  Fig.~\ref{fig10}. Extrapolation of the sequence of densities to
  higher wavelengths, especially to the wavelength of relict radiation
  photons $\lambda$=1.9 [mm] is shown. First five densities $D$ were
  calculated following the description in Section 3.2. Possible final
  error of the density $D$ is estimated and the value of the critical
  density $D_{\mathrm{critical}}$ according Smoot \& Davidson (1977) and Silk
  (1977) is given.}
\label{table:2}
\centering
\begin{tabular}{cc}
\hline\hline
Wavelength & Macroscopic density \\
$\lambda$ [nm] & $D$\ [kg.m$^{-3}$] \\
\hline
0.071069 & 108.6 \\ 
0.110674 & 40.84 \\
0.154178 & 17.18 \\
0.250466 & 4.39 \\
0.537334 & 0.46 \\[0.5ex]
\hline
\multicolumn{2}{c}{Extrapolation to higher wavelengths $\lambda$} \\
\hline
1 & 9.0 E-02 \\
10 & 6.0 E-05 \\
100 & 4.0 E-08 \\
1 000 & 2.0 E-11 \\
1 000 000 & 1.0 E-21 \\[0.5ex]
\hline
1 900 000 &  \\
= \mbox{1.9 [mm]} & {} \raisebox{1.5ex}[0pt]{\mbox{9.0 E-23$\pm$E-3}} \\[0.5ex]
\hline
\multicolumn{2}{c}{Critical density:} \\
\multicolumn{2}{c}{$D_{\mathrm{critical}}$=5.0 to 7.0 E-27 [kg.m$^{-3}$]} \\
\hline\hline
\end{tabular}
\end{table}

   \begin{figure}
   \centering
   \includegraphics[width=8cm]{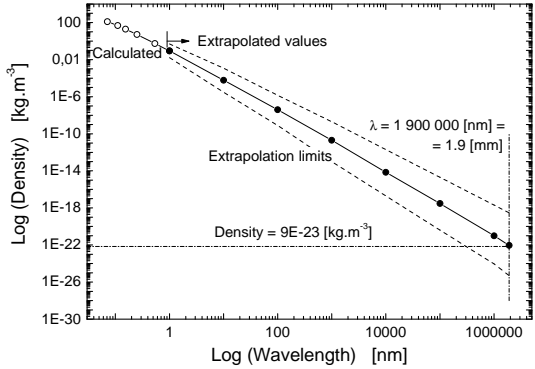}
      \caption{Extrapolation of the dependence of densities on
        wavelengths to the wavelength of relict (CMB) photons
        $\lambda$ = 1.9 [mm]. Empty circles represent values shown
        already in Fig. 11. Full circles are extrapolated
        values. Dashed lines show the limits of possible
        extrapolations.  }
         \label{fig12}
   \end{figure}


\section{Calculations in the Relict reciprocal space}

\subsection{Modelling according the Debye formula}

In the case when Fig. 4 should be an X-ray scattering picture of a
disordered material (e.g. of a glass) then such record would represent
a picture typical for a material with well developed clusters. Their
mutual distance should then characterize the position of the ``first''
massive peak. It follows from theory and experience that it is not
possible to get from this peak information on the internal structure
of Clusters, only on their magnitude and mutual distance.

The method which has to be used for an analysis of this type of
scattering is a direct calculation of scattered radiation on the basis
of the Debye formula
   
\begin{equation}
I(S_{\mathrm{Relict}}) = \sum_{i=1}^{n} \sum_{j=1}^{n} f_{i} \, f_{j} 
 \, \frac{\sin (d_{ij} S_{\mathrm{Relict}})}{d_{ij} S_{\mathrm{Relict}}}
\label{eq18}
\end{equation}

Here $f_{i}$ and $f_{j}$ are the scattering factors of $n$ input
particles and $d_{ij}$ are the distances in real space between all
available particles in the model and $S_{\mathrm{Relict}}$ is the
scattering vector in the Relict reciprocal space defined in equation
(9). It should be pointed out that as scattering factors $f_{i}$ and
$f_{j}$ we have used now the relict radiation factor
$f_{\mathrm{Relict}}$ found in Sect.\ 2.2.4. The summation is over all
$n$ particles in the model. This formula gives the average scattered
intensity for an array of particles (or atoms in solid state physics)
with a \emph{completely
random orientation in space to the incident radiation}.

   \begin{figure}
   \centering
   \includegraphics[width=8cm]{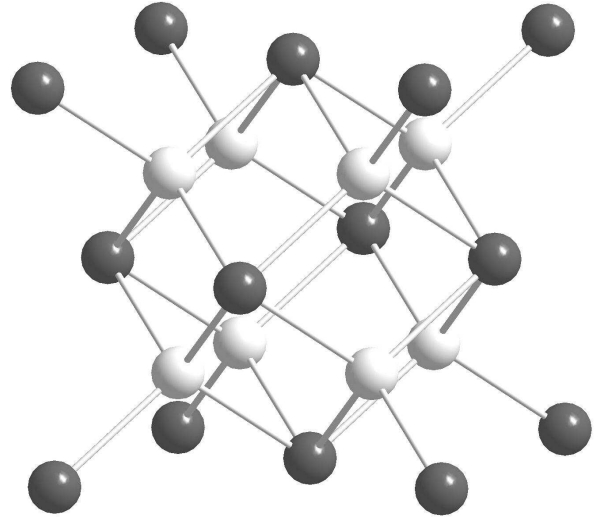}
      \caption{The basic skeleton (and-or a part of a Cluster
        structure) consists of 22 ``positions'' formed by 8 edge-bound
        tetrahedrons. All positions are identical, for a better
        graphic representation are the centres of tetrahedrons drawn
        white. The picture has been constructed using programs by by
        Pet\v{r}\'{\i}\v{c}ek (2006) and Brandenburg (1999).  }
         \label{fig13}
   \end{figure}

Our model was quite simple: For the wavelength $\lambda$ = 0.071069
[nm] the Cluster was a tetrahedron (5 particles) with an
inter-particle distance 0.263 [nm] i.e. located in a cube with an edge
0.607 [nm]. In order to find the best fit with the scattering curve
according equation (18), the distance between Clusters (tetrahedrons)
had to be $d$ = 3 [nm], i.e. the tetrahedrons were located in
positions of the basic skeleton characterized now by a side $a$ = 6.93
[nm]. This model had 22$\times$5 particles, i.e. a total of 110
particles. This calculation is shown in Fig. 14.

For all other wavelengths ($\lambda \ge$ 0.110674 [nm]) we had to
increase the dimensions of the Cluster. \emph{The Cluster had then the
  form of the skeleton shown in Fig. 13} with an edge 0.607 [nm] and
consisted of 22 particles (again with an inter-particle distance 0.263
[nm]) embedded in 8 edge-bound tetrahedrons. Only this Cluster
occupied the ``positions'' of the cubic skeleton shown in Fig. 13
forming now an Object. (A more instructive schematic presentation of
an Object is shown in Fig. 18 where Clusters are presented as small
darker circles filled with ``particles''.) When changing the dimension
of this skeleton, we simultaneously changed again the distance $d$
between Clusters. In order to reach for $\lambda$ = 0.110674 [nm] the
correct position of the massive peak at 1.6 [nm$^{-1}$] an
inter-Cluster distance $d$ = 4.65 [nm] had to be used, i.e. the
dimension of the skeleton was characterized by the side $a$ = 10.74
[nm]. This model had then 22$\times$22 particles, i.e. a total of 484
particles and simulated a part of the Object structure. The
calculation is shown in Fig. 15.

   \begin{figure}
   \centering
   \includegraphics[width=8cm]{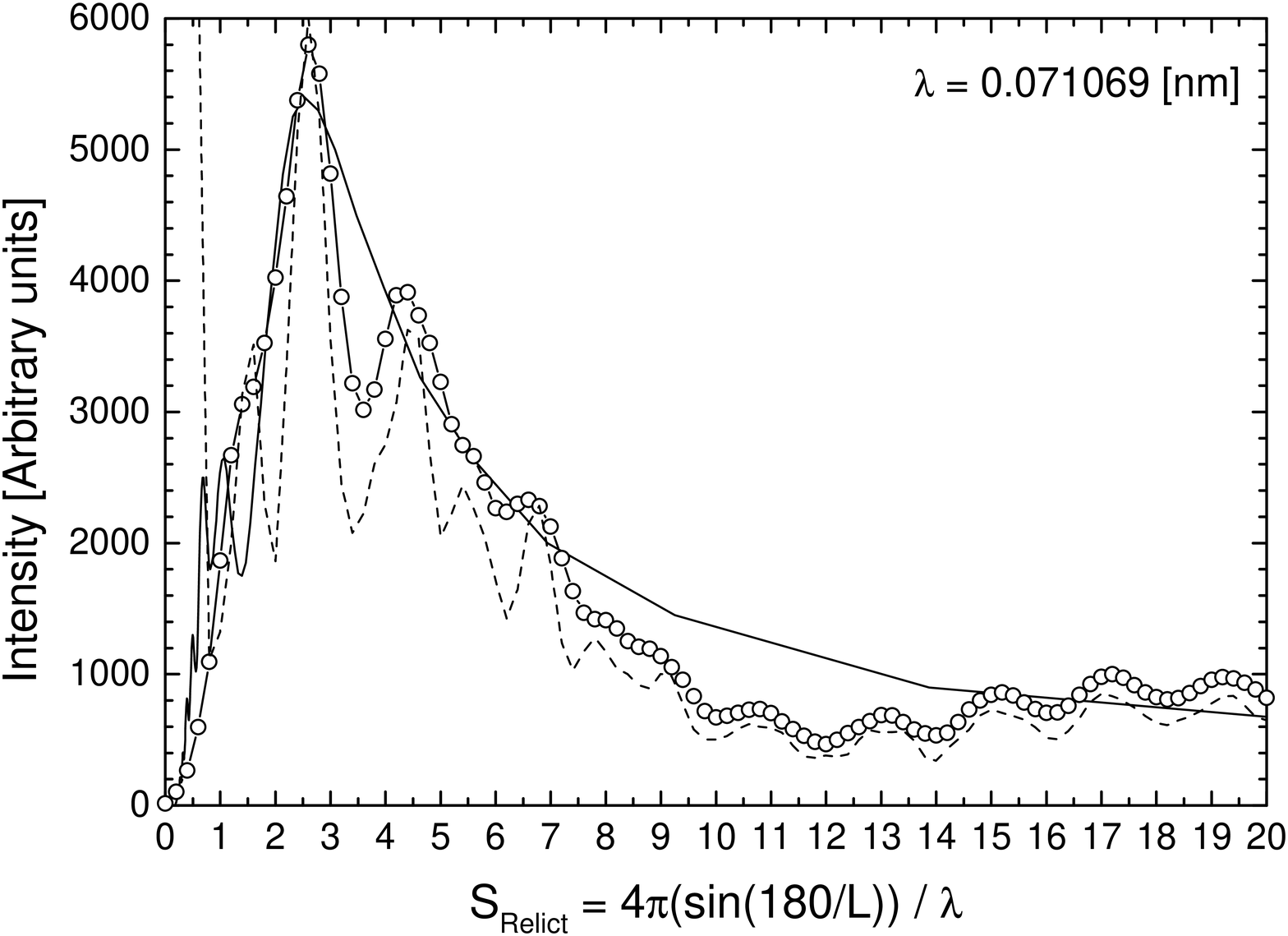}
      \caption{Calculation of the profile of the recalculated
        anisotropy spectrum for $\lambda$ = 0.071069 [nm] based on a
        set of 22 Clusters with a mutual distance $d$=3 [nm]. The
        Cluster was formed by a tetrahedron (5 particles); hence there
        were 110 particles in a model, see text for details. Full line
        - experiment, empty circles - calculated scaled and smoothed
        curve, dashed line - calculated scaled scattering.  }
         \label{fig14}
   \end{figure}

   \begin{figure}
   \centering
   \includegraphics[width=8cm]{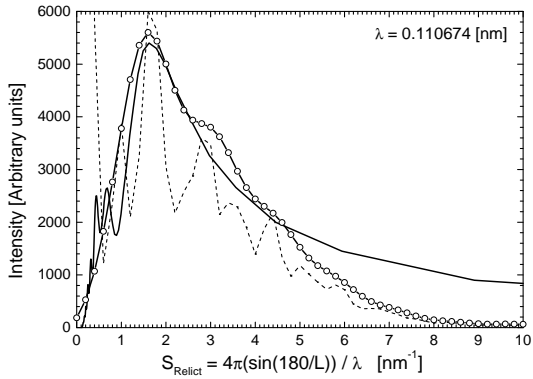}
      \caption{Calculation of the profile of the recalculated
        anisotropy spectrum for $\lambda$ = 0.110674 [nm] and for a
        set of 22 Clusters with a mutual distance of $d$=4.65 [nm]. The
        Cluster consisted of 22 particles, hence there were 484
        particles in the model, see text for details. Full line -
        experiment, empty circles - calculated scaled and smoothed
        curve, dashed line - calculated scaled scattering.  }
         \label{fig15}
   \end{figure}

   \begin{figure}
   \centering
   \includegraphics[width=8cm]{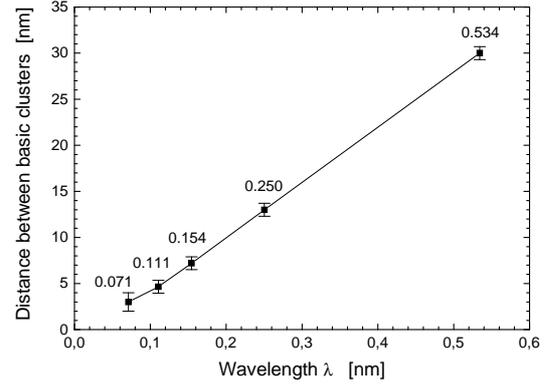}
      \caption{Dependence of distances between Clusters on the
        wavelength (numerical values are shown). With exception of the
        “0.071 case” models consisted of 22 Clusters with 22 particles
        in each Cluster, i.e. a model included 484
        particles. Inter-Cluster distances characterize the position
        of the massive peak, see Figs. 14 and 15 and the text for
        details.  }
         \label{fig16}
   \end{figure}

\begin{table}
\caption{Extrapolation of distances between Clusters to the
  wavelength of relict photons 1.9 [nm]. These distances influence the
  position of the massive peak, see Figs.~\ref{fig14} and
  \ref{fig15}. The estimate of the final error is based on errors
  given in Fig.~\ref{fig16}}
\label{table:3}
\centering
\begin{tabular}{cc}
\hline\hline
Wavelength & Distance between Clusters \\
$\lambda$ [nm] & $d$\ [nm] \\
\hline
0.071069 & 3.00 $\pm$ 1.50 \\ 
0.110674 & 4.65 $\pm$ 1.00 \\
0.154178 & 7.20 $\pm$ 1.00 \\
0.250466 & 13.00 $\pm$ 1.00 \\
0.537334 & 30.00 $\pm$ 1.00 \\[0.5ex]
\hline
\multicolumn{2}{c}{Extrapolation to higher wavelengths $\lambda$} \\
\hline
1 & 60.8 \\
10 & 608 \\
100 & 6 081 \\
500 & 30 404 \\
1 000 & 60 808 \\
1 000 000 & 60 807 919 \\
1 900 000 &  115 535 046 \\[0.5ex]
\hline
= 1.9 [mm] & $d_{\mathrm{Relict}}$=12$\pm$1 [cm] \\
\hline\hline
\end{tabular}
\end{table}

   \begin{figure}
   \centering
   \includegraphics[width=8cm]{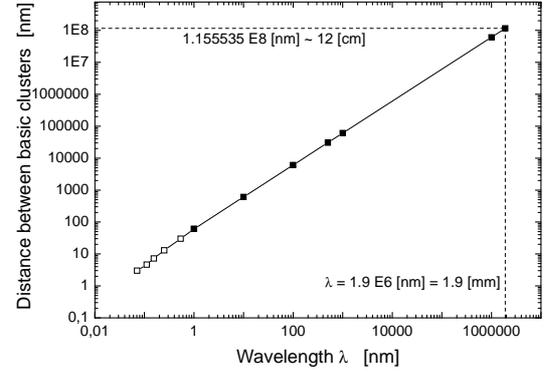}
      \caption{Extrapolation of distances between Clusters to the
        wavelength 1.9 [nm] - full squares; calculated values - empty
        squares (see Fig. 16 and Table 3).  }
         \label{fig17}
   \end{figure}

Calculations of Cluster distances for additional wavelengths
(0.154178, 0.250466 and 0.537334 [nm]) have shown (see Fig. 16) that
the dependence of Cluster distances on the corresponding wavelength is
linear. This fact enabled an extrapolation of the Cluster distance
$d$ to the wavelength of relict photons
$\lambda$=1.9 [nm], see Table 3. This extrapolated distance is
$d_{\mathrm{Relict}}$= (12$\pm$1)[cm].  The extrapolation is
visualized in Fig. 17.

It should be noted that the recalculated anisotropy spectrum depends
in this case directly on the angle $\theta_{\mathrm{Relict}}$ which is
equal to the angle $\alpha$ (see equation (8)) and therefore a
recalculation of the inter-Cluster distance $d$ into a real space
distance is not necessary because the Debye formula analyzes the
Relict reciprocal space represented by the vector
$S_{\mathrm{Relict}}$ directly in real space distances, see the
quantity $d_{ij}$ in equation (18).


\subsection{Quantitative relations between Objects, Clusters and
  particles} 

\subsubsection{Estimates from the Object distances}

We have found that the nearest distance between Objects (big clusters)
is $\sim98$ [m], see Table 1. In this moment we suppose a relatively
simple organization of Objects, i.e. a ``cubic body-centred''
arrangement, in which an Object in the centre has 8 nearest neighbour
Objects distant $b_{O}$ = 98 [m], where $b_{O}$ is the half of the
body diagonal in a cube with a side
\begin{equation}
a_{O} = (2b_{O}) / \sqrt{3} = (2\times98 \mathrm{[m]}) / 1.732 =
113.164 \mathrm{[m]}. 
\label{eq19}
\end{equation}

The volume $V_{O}$ of this cube is therefore
\begin{equation}
V_{O} = 1 449 188 \mathrm{[m^{3}]} = 1.449 \times 10^{6} \mathrm{[m^{3}]}.
\label{eq20}
\end{equation}

Using now our result on the density of the matter, see Table 2,
\begin{equation}
D = 9 \times 10^{-23}\mathrm{[kg.m^{-3}]} = m_{2O}/V_{O} 
\label{eq21}
\end{equation}

we are able to calculate in this model the mass $m_{2O}$ of Objects
embedded in a cube with the volume $V_{O}$.
\begin{eqnarray}
m_{2O} & = & D \times V_{O} = 9 \times 10^{-23}\mathrm{[kg.m^{-3}]} \times 1.449
\times 10^{6} \mathrm{[m^{3}]} \nonumber \\ 
 & = & 13.04 \times 
10^{-17} \mathrm{[kg]}.      
\label{eq22}
\end{eqnarray}

At the same time, however, we have to take in account that, as a
matter of fact, there are two Objects in the space of
the cube (in each cube corner there is only 1/8 of the second
Object). Hence \emph{the mass} $m_{O}$ \emph{embedded in one Object} is
\begin{equation}
m_{O} = 6.52 \times 10^{-17} \mathrm{[kg]}.  
\label{eq23}
\end{equation}

{\itshape A) The mass is formed by a 1:1:1 mixture of protons, helium
  nuclei and electrons }

We may suppose now that the universe (in the time when the microwave
background radiation began propagating) consisted of baryons
(protons, helium nuclei, etc) and electrons, neutrinos, photons and
dark matter particles. Supposing now that we have a mixture consisting
of protons, helium nuclei and electrons in a relation 1:1:1, then the
\emph{medium mass of a ``particle''} $m_{\mathrm{part}}^{1:1:1}$ in
this mixture is
\begin{eqnarray}
m_{\mathrm{part}}^{1:1:1} & = & \{(1.67 \times10^{-27} \mathrm{[kg]}) + (6.64
  \times10^{-27} \mathrm{[kg]}) \nonumber \\ 
 & & + (0.00091 
  \times10^{-27} \mathrm{[kg]})\}/3,  \quad\mbox{i.e.} \nonumber \\
m_{\mathrm{part}}^{1:1:1} & = & (8.311 / 3) \times10^{-27} \mathrm{[kg]} = 2.77
\times10^{-27} \mathrm{[kg]}  
\label{eq24}
\end{eqnarray}

and the \emph{number of particles} $N_{\mathrm{partO}}^{1:1:1}$
\emph{in one Object} is in this case
\begin{eqnarray}
N_{\mathrm{partO}}^{1:1:1} & = & m_{O}/ m_{\mathrm{part}}^{1:1:1} \nonumber \\ 
 & = & (6.52 \times10^{-17}
\mathrm{[kg]}) / (2.77 \times10^{-27} \mathrm{[kg]}), \quad\mbox{i.e.} 
\nonumber \\ 
N_{\mathrm{partO}}^{1:1:1} & = & 2.35 \times 10^{10} \quad\mbox{particles}
\label{eq25}
\end{eqnarray}

{\itshape B) The mass is formed by a 1:1:10 mixture of protons, helium
  nuclei and electrons }

Supposing now a mixture consisting of protons, helium nuclei and
electrons in a relation 1:1:10, then \emph{the medium mass of a
  ``particle''} $m_{\mathrm{part}}^{1:1:10}$ in this system is
\begin{eqnarray}
m_{\mathrm{part}}^{1:1:10} & = & \{(1.67 \times10^{-27} \mathrm{[kg]}) + (6.64
  \times10^{-27} \mathrm{[kg]}) + \nonumber \\ 
 & & 10 \times(0.00091 \times10^{-27}
  \mathrm{[kg]})\}/12, \nonumber \quad\mbox{i.e.} \\  
m_{\mathrm{part}}^{1:1:10} & = & (8.319 / 12) \times10^{-27}
\mathrm{[kg]} = 0.69 \times10^{-27} \mathrm{[kg]}   
\label{eq26}
\end{eqnarray}

and the \emph{number of particles} $N_{\mathrm{partO}}^{1:1:10}$
\emph{in one Object} is then
\begin{eqnarray}
N_{\mathrm{partO}}^{1:1:10} & = & m_{O}/m_{\mathrm{part}}^{1:1:10} \nonumber \\ 
 & = & (6.52 \times10^{-17}
\mathrm{[kg]}) / (0.69 \times10^{-27} \mathrm{[kg]}), \nonumber
\quad\mbox{i.e.} \\
N_{\mathrm{partO}}^{1:1:10} & = & 9.44 \times 10^{10} \quad\mbox{particles}.
\label{eq27}
\end{eqnarray}

This section may be summarized by the statement that there are
\begin{equation}
N_{\mathrm{partO}} \approx 10^{11}\quad\mbox{particles in one Object}.  
\label{eq28}
\end{equation}


\subsubsection{Estimates from Cluster distances}

According our calculations the distance between Clusters is $\sim$12
[cm] = $1.2\times10^{-1}$ [m], see Table 3 and Fig. 18. Similarly as
in the previous case we suppose again a relatively simple organization
of Clusters, i.e. a cubic body-centred arrangement in which a Cluster
in the centre has 8 “nearest neighbour” Clusters distant $b_{C} = 1.2
\times 10^{-1}$ [m], where $b_{C}$ is the half of the body diagonal in
a cube with a side
\begin{equation}
a_{C} = (2b_{C}) / √3 = (2.4 \times 10^{-1} \mathrm{[m]}) / 1.732 =
1.386 \times 10^{-1} \mathrm{[m]}.  
\label{eq29}
\end{equation}

The volume $V_{C}$ of this cube is therefore
\begin{equation}
V_{C} = 2.66 \times 10^{-3} \mathrm{[m^{3}]}. 
\label{eq30}
\end{equation}

Using now our result on the density of the matter, see already equation (21)
\begin{displaymath}
D = 9 \times 10^{-23}\mathrm{[kg.m^{-3}]} = m_{2C}/V_{C} ,
\end{displaymath}

we are able to calculate for this model the mass $m_{2C}$ of Clusters
embedded in a cube having the volume $V_{C}$,
\begin{eqnarray}
m_{2C} & = & D \times V_{C} = 9 \times 10^{-23}\mathrm{[kg.m^{-3}]} \times 2.66
\times 10^{-3} \mathrm{[m^{3}]} \nonumber \\ 
 & = & 23.94 \times 10^{-26}
\mathrm{[kg]}. 
\label{eq31}
\end{eqnarray}

Here again we have to take in account that there are two Clusters in
the space of the cube (in each corner there is only 1/8 of the second
Cluster). Hence \emph{the mass} $m_{C}$ \emph{embedded in one Cluster
  is}
\begin{equation}
m_{C} = 1.20 \times 10^{-25} \mathrm{[kg]}. 
\label{eq32}
\end{equation}

{\itshape A) The mass is formed by a 1:1:1 mixture of protons, helium
  nuclei and electrons }

Similarly as in the preceding Sect.\ 4.2.1. we suppose again a mixture
of protons, helium nuclei and electrons in a relation 1:1:1,
respectively. The medium mass of a ``particle''
$m_{\mathrm{part}}^{1:1:1}$ in this mixture is (see equation (24))
\begin{displaymath}
m_{\mathrm{part}}^{1:1:1} = 2.77 \times 10^{-27} \mathrm{[kg]}
\end{displaymath}

and \emph{the number of particles} $N_{\mathrm{partC}}^{1:1:1}$ \emph{in one Cluster}
is then
\begin{eqnarray}
N_{\mathrm{partC}}^{1:1:1} & = & m_{C}/m_{\mathrm{part}}^{1:1:1} \nonumber \\ 
 & = & (1.20 \times10^{-25}
\mathrm{[kg]}) / (2.77 \times10^{-27} \mathrm{[kg]}), \nonumber
\quad\mbox{i.e.} \\
N_{\mathrm{partC}}^{1:1:1} & = & 0.43 \times 10^{2}\quad\mbox{particles}. 
\label{eq33}
\end{eqnarray}

{\itshape B) The mass is formed by a 1:1:10 mixture of protons, helium
  nuclei and electrons }

Identically as in the preceding Sect.\ 4.2.1. the medium mass of a
``particle'' is in this case, see equation (26),
\begin{displaymath}
m_{\mathrm{part}}^{1:1:10} = 0.69 \times10^{-27} \mathrm{[kg]}
\end{displaymath}

and the \emph{number of particles} $N_{\mathrm{partC}}^{1:1:10}$ \emph{in
one Cluster} is then
\begin{eqnarray}
N_{\mathrm{partC}}^{1:1:10} & = & m_{C}/ m_{\mathrm{part}}^{1:1:10} \nonumber \\ 
 & = & (1.20 \times10^{-25}
\mathrm{[kg]}) / (0.69 \times10^{-27} \mathrm{[kg]}), \nonumber
\quad\mbox{i.e.} \\
N_{\mathrm{partC}}^{1:1:10} & = & 1.74 \times 10^{2} \quad\mbox{particles}.  
\label{eq34}
\end{eqnarray}

This section can be summarized by the statement that there are
\begin{equation}
N_{\mathrm{partC}} \approx 10^{2} \quad\mbox{particles in one Cluster}.
\label{eq35}
\end{equation}


\subsubsection{Consequences of previous calculations}

We are now able to calculate easily the number of Clusters in one
Object. Because an Object consists of $N_{\mathrm{partO}} \approx
10^{11}$ particles in one Object (equation (28)) and there are
$N_{partC} \approx 10^{2}$ particles in one Cluster, it follows that
an Object should be composed from $N_{C}$ Clusters, where
\begin{equation}
N_{C} = (N_{\mathrm{partO}} / N_{partC}) \approx 10^{11}/10^{2} \approx 10^{9}
\quad\mbox{Clusters}.  
\label{eq36}
\end{equation}

Supposing that densities in the Object and in the Cluster are equal
then this value is independent on the value of the density and on the
mass of the particle (e.g. $m_{\mathrm{part}}^{1:1:1}$) and depends
only on the relation of the volumes $V_{O}/V_{C}$, because
\begin{eqnarray}
N_{C} & = & (N_{\mathrm{partO}} / N_{partC}) = V_{O}/V_{C} \nonumber \\ 
 & = & 1.449\times10^{6}
\mathrm{[m^{3}]} / 2.66\times10^{-3} \mathrm{[m^{3}]} \nonumber \\
& \approx & 10^{9}\quad\mbox{Clusters}. 
\label{eq37}
\end{eqnarray}


\section{Discussion}

In the following discussion we will concentrate on several important
ideas which may arise when reading this paper.

First of all this contribution should demonstrate how the formalism
imported from solid state physics could be useful in solving specific
cosmological problems: It may shed some new light on the physical
processes taking place in the primordial plasma.

For example, this work points quite unequivocally to clustering
processes and to a cluster-like structure of the matter in the moment
when the universe became transparent for photons (see Sect.\ 5.1.).

Further, the new formalism enabled us a simple and general description
of the interaction of relict radiation with the matter and may help in
an improvement of the theoretical predictions of the CMB pattern (see
Sect.\ 5.2.).

Finally this new approach may be useful in the analysis of the CMB
data. We have shown that the transformation of the anisotropy spectrum
of relict radiation into a special two-fold reciprocal space and into
a simple reciprocal space was able to bring quantitative data in real
space. Problems with the transformation into reciprocal spaces, mainly
with the use of the proper wavelength of relict photons will be
discussed in Sect.\ 5.3.


\subsection{The cluster-like structure of the primordial matter}

   \begin{figure}
   \centering
   \includegraphics[width=8cm]{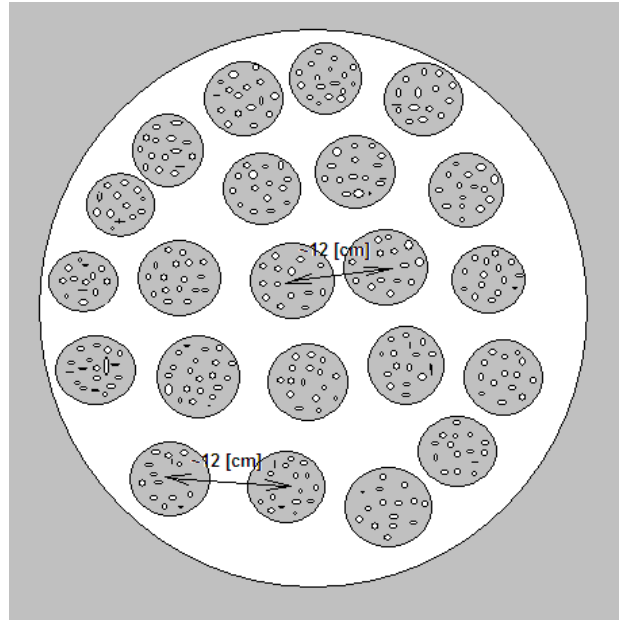}
      \caption{A schematic arrangement of Clusters (darker regions)
        with particles (small white points) in an Object (white
        region). Distance between Objects is $\sim$98 [m], see Table
        1. Detailed structure of a Cluster and of an Object in our
        model is presented in Fig. 13. The most probable model
        distance between Clusters is 12 [cm], see Table
        3. Distance between particles is 0.26 [nm]. There are
        $\sim$10$^{11}$ particles in one Object, $\sim$10$^{2}$
        particles in one Cluster (see Sect.\ 4.2.) and $\sim$10$^{9}$
        Clusters in one Object, see Sect.\ 4.2.  }
         \label{fig18}
   \end{figure}

We have already mentioned that the process of forming the primordial
matter by particle clustering may be a new physical effect which has
not been fully taken into consideration in the past. Now we present a
model of the cluster-like structure

Concerning our results on the distances between Objects and Clusters,
we have arrived to three numbers, which we interpret in a following
way: The first one, which is 98 [m], (Table 1) indicates the distance
between Objects (big clusters), the second one, which is 12 [cm],
(Table 3) indicates the distance between smaller Clusters, while the
internal structure of a single Cluster is formed by 22 particles and is
characterized by a medium particle distance 0.26 [nm], see Sect.\ 4.1.

In Fig.18 we show a schematic picture of the cluster model. The big
circle represents an Object. An Object is a clump of Clusters, where
only a part of this clump was simulated in our model by 22 Clusters
each having 22 particles, i.e. by a total of 484 particles in an
Object.

Although this model gave a sufficiently well agreement with the width
of the massive peak, as demonstrated in Fig. 15, our estimates
(Sect.\ 4.2.) show that the number of Clusters as well as the number of
particles in one Cluster is greater., i.e. that there may be as far as
10$^{11}$ particles in one Object and 10$^{2}$ particles in one Cluster. That
the density plays an important role in these calculations will be
discussed in Sect.\ 5.4.

It is important to note that the distance between Objects
($^{2}R$=98[m], see Table 1) is not identical with the dimension of
the Object as defined in Sect.\ 4.1. There the dimension of an Object
was determined by the inter-Cluster distance $d_{\mathrm{Cluster}}$ =
0.12 [m] (see Table 3). This distance is a quarter of the body
diagonal in the cube-like skeleton (with an edge $a_{\mathrm{Object}}$
= 0.28 [m]) simulating the Object according Fig. 13. The dimension of
an Object is then determined by the diameter of a sphere surrounding
Clusters located in the skeleton ``positions''. The value of this
diameter is 2$R_{\mathrm{Object}}$ = 0.48 [m], i.e. much
smaller than $^{2}R$=98[m].

We have already mentioned (see Sect.\ 4.1.) that in principle it is
not possible to solve on the basis of the massive peak (located
e.g. at $S_{\mathrm{Relict}}$ = 1.62 [nm$^{-1}$] for $\lambda$ =
0.1107 [nm], see Fig. 15) the internal structure of a Cluster. It is
possible to reach only information on the Cluster magnitude and on the
distance between Clusters.

Just this information we have derived from our model calculations: The
magnitude of a Cluster was based on the particle distance 0.263 [nm]
and was defined by a cube with an edge $a_{\mathrm{Cluster}}$ = 0.607
[nm] (see Fig. 13), which may be surrounded by a sphere with a
radius $R_{\mathrm{Cluster}}$ = 0.53 [nm], hence a diameter of a
Cluster has the value 2$R_{\mathrm{Cluster}}$ = 1.05 [nm].  It is
important to note that this
diameter, similarly as for Objects, is not identical with the distance
between Clusters (12 [cm]), see Fig. 18. However, it is this distance,
which influences the position of the peak, while its
intensity depends on the number of particles in the Cluster and their
``mass'' represented by their ``scattering'' power (i.e. by their
relict radiation factor).

Interesting may be the effect of changing the inter-Cluster distance:
A decrease of the inter-Cluster distance from 12 cm to e.g. 10 cm
would bring for the 1:1:1 mixture of particles (see Sect.\ 4.2.2.) a
value of 25 particles in one Cluster and for the 1:1:10 mixture of
particles a value of 100 particles in a Cluster, i.e. numbers which
roughly correspond to our model numbers in Sect.\ 4.1. Similarly a
greater proportion of heavier particles should decrease the number of
particles, thus again corresponding to the model number.

Even when the cluster model gave a good profile of the massive peak at
e.g. 1.62 [nm], than such a model cannot be a unique one, because the
calculation of the profile is not sensitive to the internal cluster
structure, however, the cluster-like character of the modelling
process has to be maintained.

How it was possible to estimate on the basis of inter-Cluster and
inter-Object distances the number of particles (protons, helium
nuclei, electrons) in Objects and Clusters and of Clusters in an
Object was demonstrated in Sects.\ 4.2. and 4.3., however how these
numbers are influenced by the density will be discussed in Sect.\ 5.4.


\subsection{The relict radiation factor}

We have already pointed out in Sect.\ 2.1.1. why during the analysis
of the CMB spectrum it has not been possible to apply conventional
atomic scattering factors used in solid state physics and why a new
special factor reflecting the complexity of interaction processes of
photons with the primordial matter has to be constructed. It is
important to have in mind that the description of these interactions
is possible \emph{only in a special two-fold reciprocal space} into
which the CMB spectrum was transformed. We have called this new factor
the \emph{relict radiation factor} and it had to substitute all
complicated processes which participate in the formation of the
angular power spectrum of CMB radiation, see Sect.\ 2.2.

Because relict photons realize their interaction with various kinds of
particles and we have generated only one radiation factor, this factor
represents, as a matter of fact, a medium from all possible individual
relict radiation factors. In this way \emph{this new formalism offers a
general description of the interaction of relict radiation with the
matter} and simultaneously reflects the complexity of processes which
influence the anisotropy spectrum of CMB radiation from the
cosmological point of view (Hu et al. 1995).

During our study we have concentrated on three important facts which
may justify the attempt to interpret the anisotropy spectrum of CMB
radiation as a consequence of the interaction of photons with density
fluctuations characterizing the distribution of particles before the
recombination process.

The first fact is that temperature fluctuations in the CMB spectrum
are related to fluctuations in the density of matter in the early
universe and thus carry information about the initial conditions for
the formation of cosmic structures such as galaxies, clusters or voids
(Wright 1994).

Secondly, it is the fact that the information on these density
fluctuations in the distribution of particles (electrons, ions, etc.)
has been brought by photons. Photons which we observe from the
microwave background have traveled freely since the matter was highly
ionized and they realized their last Thomson scattering (see already
Sect.\ 2.1.1.). If there has been no significant early heat input from
galaxy formation then this happened when the Universe became cool
enough for the protons to capture electrons, i.e. when the
recombination process started (White 1994).

The third fact is that the anisotropy spectrum is angular dependent,
see Fig. 1.

Although we know that the anisotropy spectrum of CMB radiation, as
presented in Fig. 1, has no direct connection with a scattering
process of photons, it was the transformation of the CMB spectrum into
a two-fold reciprocal space, which enabled us to interpret the
anisotropy spectrum of CMB radiation as a \emph{result of an
  interaction process of photons with density fluctuations} of the
matter represented by electrons, ions or other particles. This
approach enabled us to reach an advantageous approximation of this
process.

The process consisted of two steps: First of all we have constructed
in Sect.\ 2.2.1. an angular reciprocal space characterized by the
``scattering'' angle $\theta_{\mathrm{Classic}}$, see equations (2)
and (4).  This space is reciprocal to the space characterized by the
angle $\alpha$ ($\alpha$ is the angle between two points in which
temperature fluctuations of CMB radiation are compared to an overall
medium temperature).

Then, we have constructed an additional ``classic'' reciprocal space
(1/$\lambda$) into which the first one (the
$\theta_{\mathrm{Classic}}$, space) was dipped, by defining in this
new ``two-fold'' reciprocal space the classic scattering vector
$s_{\mathrm{Classic}}$, see equation (6). Only after these
transformations we treated in this new Classic reciprocal space the
transformed anisotropy CMB spectrum as a scattering picture of relict
photons.

It was only this space in which we simulated (in Sect.\ 2.2.4.) the
interaction of CMB (relict) photons with density fluctuations by the
relict radiation factor $f_{\mathrm{Relict}}$.

The criterion for the trial and error construction of the relict
radiation factor $f_{\mathrm{Relict}}$ has been that this factor had
to fulfill the three requirements set at the beginning of
Sect.\ 2.2.4. Only then it was secured that after the Fourier
transform, according equations (A.2) and-or (15), there will not be
any (or at least small) parasitic fluctuations on the curve $\rho(r)$
and-or $\rho^{\mathrm{Fourier}}(r)$.  That we have achieved these
demands is documented in Fig. 8 where we do not see any parasitic
fluctuations on the curve $\rho^{\mathrm{Fourier}}(r)$ and
as a consequence on the curve $\rho(r)$.

To summarize: It is true that in our formal analogy between scattering
of e.g. short-wave radiation on disordered matter (Fig. 2) and
``scattering'' of CMB photons on electrons, ions and other particles
(Fig. 1) is an essential difference, because the physical processes
are completely different, e.g. the scattering process itself, length
scales involved, etc., however, the difference between physical
processes is reflected and simultaneously eliminated by the special
relict radiation factor $f_{\mathrm{Relict}}$ (Sect.\ 2.2.4.), which
we have included into all calculations based on the classic two-fold
\emph{reciprocal space} (see Sect.\ 2.1.). Moreover, additional
calculations in the Relict reciprocal space (see Sect.\ 4.) based on
the relict radiation factor were done directly for the
\emph{transformed} angular power spectrum of relict radiation (see
$I_{\mathrm{Relict}}(S_{\mathrm{Relict}})$ in Fig. 4) and thus present
an information on distance relations between Clusters (formed by
particles) in \emph{real space}.


\subsection{The wavelength problem}

The problem is to which wavelength of relict photons we have to relate
our calculations. One possibility may be to refer this wavelength to
that time when 379.000 years after the Big Bang the Universe cooled
down to 3000 K and the ionization of atoms decreased already only to
1\%. Then according Wien's law   
\begin{equation}
\lambda_{\mathrm{max}} = \frac{b}{T}
\label{eq38}
\end{equation}

where $\lambda_{\mathrm{max}}$ is the peak wavelength, $T$ is the
absolute temperature of the blackbody, and $b$ is a constant of
proportionality called Wien's displacement constant, $b$ =
2.8978$\times$10$^{-3}$ [mK], we obtain for the temperature 3000 K a
wavelength value $\lambda_{\mathrm{max}}$ = 966 [nm] (\v{S}m\'{\i}da
2010).

However, simultaneously we must be aware of the fact that we are
analyzing CMB photons \emph{now} when the temperature of the universe, due to
its expansion, is 2.725 K. Then the wavelength of photons according
the Wien's law should be $\sim$1 [mm].

On the other hand the COsmic Background Explorer (COBE) measured with
the Far Infrared Absolute Spectrophotometer (FIRAS) the frequency
spectrum of the CMB, which is very close to a blackbody with a
temperature 2.725 K (Mather et al. 1994; Wright et al. 1994). The
results are shown in Fig. 19 in units of intensity (see the text to
Fig. 19). It follows that the wavelength corresponding to the maximum
is 1.9 [mm].

   \begin{figure}
   \centering
   \includegraphics[width=8cm]{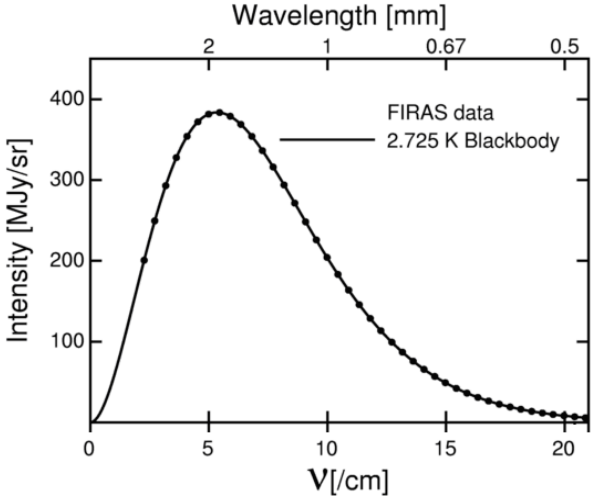}
      \caption{Dependence of the intensity of the CMB radiation on
        frequency as measured by the COBE Far InfraRed Absolute
        Spectrophotometer (FIRAS) (Mather et al. 1994; Wright et
        al. 1994). The thick curve is the experimental result; the
        points are theoretically calculated for an absolute black body
        with a temperature of 2.725 [K]. The x axis variable is the
        frequency $\nu$ in [cm$^{-1}$]. The y-axis variable is the
        power per unit area per unit frequency per unit solid angle in
        MegaJanskies per steradian (1 [Jansky] is a unit of
        measurement of flux density used in radioastronomy,
        abbreviated ``Jy'' (1 [Jansky] is 10$^{-26}$
        [W.m$^{-2}$.Hz$^{-1}$]).  }
         \label{fig19}
   \end{figure}

After all we have decided to relate our results to the wavelength of
CMB photons 1.9 [mm] which corresponds to the maximum of the intensity
distribution. Because the distribution of the spectrum covers a
relatively broad interval of wavelengths, see Fig. 19, calculations
based on the wavelength 1.9 [mm] should then represent the most
probable calculation and estimate presented in this study. Moreover,
this consideration is supported by the fact that the angular
distribution of CMB radiation is the same for all wavelengths.

However, on the basis of graphs in Figs. 10, 12 and 16 an easy
recalculation of distances and-or of the density would be possible
when another CMB photons wavelength would be considered as more
appropriate.


\subsection{The density of the mass and distances between Objects,
  Clusters and particles} 

The way how we arrived to numbers characterizing the density of the
matter was described in Sect.\ 3.2. In a conventional X-ray analysis
the density is the macroscopic density of the material under
study. Therefore we suppose that also in this case the density which
influences the parabolic shape of the curve of total disorder (see the
first member on the right side of equation (A.2) and-or (15) and
Fig.8) should be understood as a \emph{real medium density of density
  fluctuations}.

The dependence of the density on the wavelength as demonstrated in
Figs. 11 and 12 is not perfectly linear; therefore we have marked in
Fig. 12 the extent of possible linear dependences. This result can be
formally written as
\begin{equation}
D = 10^{-22} \pm 10^{-3} \mathrm{[kg.m^{-3}]}. 
\label{eq39}
\end{equation}

It follows that this medium value is about 10$^{5}$
times higher than the ``critical density''
$D_{\mathrm{critical}}$ = (5 to 7)$\times$10$^{-27}$ [kg.m$^{-3}$]
(Smoot \& Davidson (1977), Silk (1977)), see Table 2. 

In this connection a remark should be added on the influence of the
density on the calculated numbers of particles in Objects and Clusters
(see Sects.\ 4.2.1. and 4.2.2.). Having in mind the value of the
density (expression (39)) and repeating the calculations in these
sections for the upper and lower density limit, we will receive the
number of particles in an Object in the range from 10$^{8}$ to
10$^{14}$ and the number of particles in a Cluster in the range from 0
to 10$^{5}$ particles. Because a Cluster cannot be ``empty'', the
latter numbers indicate that the lower density limit should be higher
and could reach a more probable value of $\sim10^{-23}$
[kg.m$^{-3}$]. Hence the value of the density may be then formally
written as $D$=10$^{-22}\pm10^{-1}$ [kg.m$^{-3}$].

Further, we should have in mind that the local density in a Cluster or
in an Object has to be greater. We are able to document this fact on
the basis of our Cluster model. Based on particle distances
$d_{particles}$ = 0.263 [nm], we have simulated a part of the Cluster
structure by a cube with an edge $a_{\mathrm{Cluster}}$ = 0.607
[nm]. There were 22 particles in this cube which can be closed in a
sphere with a radius $R_{\mathrm{Cluster}}$ = 2$d_{particles}$ = 0.53
[nm]. The volume of this sphere is $V_{\mathrm{Cluster}}$ = 0.62
[nm$^{3}$] = 0.62$\times$10$^{-27}$[m$^{3}$].  Supposing that
particles are represented according expression (24) by their medium
mass $m_{\mathrm{part}}^{1:1:1}$ = 2.77$\times$10$^{-27}$ [kg], we
obtain for the {\em density of the Cluster} the value
\begin{eqnarray}
D_{\mathrm{Cluster}} & = & (22\times m_{\mathrm{part}}^{1:1:1}) /
V_{\mathrm{Cluster}} = 60.94 / 0.62 
\nonumber \\ 
 & = & 98 \mathrm{[kg.m^{-3}]}, 
\label{eq40}
\end{eqnarray}

i.e. a value approaching density values known from solid state physics
(i.e. values lying between the densities of gases and liquids).

In a similar way it is possible to calculate the density in an
Object. In our model, according Table 3, the distance between Clusters
describing a part of the Object structure (corresponding the
wavelength $\lambda$=1.9 [nm]) was $d_{\mathrm{Cluster}}$=0.12
[m]. The skeleton simulating the Object had an edge
$a_{\mathrm{Object}}$=0.28 [m] and could be surrounded by a sphere
with a diameter $R$ = 2$d_{\mathrm{Cluster}}$ = 0.24 [m] and a volume
$V_{\mathrm{Object}}$ = 0.058 [m$^{3}$]. Using again the medium mass
of particles according expression (24) $m_{\mathrm{part}}^{1:1:1}$ =
2.77$\times$10$^{-27}$ [kg] and taking in account that there are
according expression (25) $N_{\mathrm{partO}}^{1:1:1}$ =
2.35$\times$10$^{10}$ particles in the Object, then the total mass in
the Object is 6.51$\times$10$^{-7}$ [kg] and we obtain for the
\emph{density of an Object} the value
\begin{eqnarray}
D_{\mathrm{Object}} & = & (N_{\mathrm{partO}}^{1:1:1} \times
m_{\mathrm{part}}^{1:1:1}) / 
V_{\mathrm{Object}} \nonumber \\ 
 & = & (6.51\times10^{-7} \mathrm{[kg]}) / 0.058 \mathrm{[m^{3}]},
\nonumber 
\quad\mbox{i.e.} \\ 
D_{\mathrm{Object}} & = & 2.24 \times 10^{-5} \mathrm{[kg.m^{-3}]},   
\label{eq41}
\end{eqnarray}

i.e. a value of density by an order $\sim$10$^{18}$ greater than the
value of the medium density of the matter $D$ = 9$\times$10$^{-23}$
[kg.m$^{-3}$] as found from the RDF analysis (see Table 2). This is a
reasonable result because there has to be a non zero value of density
in the inter-Object space.

At the same time we have to take in account that the estimates
concerning the density of matter are really complicated. The microwave
light seen by the Wilkinson Microwave Anisotropy Probe (WMAP),
suggests that fully 72\% of the matter density in the universe appears
to be in the form of dark energy (Wheeler 2007) and 23\% is dark
matter. Only 4.6\% is ordinary matter. So less than 1 part in 20 is
made out of matter we have observed experimentally or described in the
standard model of particle physics. Of the other 96\%, apart from the
properties just mentioned, we know ``absolutely nothing'' (Smolin
2007)].  In this connection we consider the density value we have
  received (9$\times$10$^{-23}$ [kg.m$^{-3}$]) as the density of the
  ordinary matter.

Last remark should be given to the probability of Object interactions
in the case of their apparently large mutual distances ($\sim$10$^{2}$ [m]). It
follows from the Maxwell speed distribution that the root mean square
particle velocity $\nu$ corresponding to the temperature $T$ = 3000 [K], is
\begin{equation}
\nu = \sqrt{\frac{3kT}{m}}
\label{eq42}
\end{equation}

where $k$ is the Boltzmann constant ($k$ = 5.4$\times$10$^{-23}$
[Joule.K$^{-1}$]) and $m$ is the mass of the particle, which may be
here the already mentioned mass of the proton ($m$ =
1.67$\times$10$^{-27}$ [kg]).  Then we obtain $\nu$ =
$\sqrt{(3\times1.38\times10^{-23} \times 3\times10^{3}) /
  (1.67\times10^{-27})}$ =
$\sqrt{(12.42\times10^{-20}/1.67\times10^{-27}}$ = 8.6$\times$10$^{3}$
[m.s$^{-1}$]. This is already a velocity, which should make possible an
intensive interaction of Objects formed by Clusters consisting of
particles.


\section{Conclusions}

A formalism of solid state physics has been applied to provide an
additional tool for the research of cosmological problems. It was
demonstrated how this new approach could be useful in the analysis of
the CMB data. After a transformation of the anisotropy spectrum of
relict radiation into a special two-fold reciprocal space it was
possible to propose a simple and general description of the
interaction of relict photons with the matter 380.000 years after the
Big-Bang by a ``relict radiation factor''. This factor, which may help
in an improvement of the theoretical predictions of the CMB pattern,
enabled us to process the transformed CMB anisotropy spectrum by a
Fourier transform and thus arrive to a radial electron density
distribution function (RDF) in a reciprocal space.

As a consequence it was possible to estimate distances between Objects
of the order of $\sim$100 [m] and the density of the ordinary matter
$\sim$10$^{-22}$ [kg.m$^{-3}$]. Another analysis based on a
direct calculation of the CMB radiation spectrum after its
transformation into a simple reciprocal space and combined with
appropriate structure modelling confirmed the cluster structure. It
indicated that the internal structure of Objects may be formed by
Clusters distant $\sim$10 [cm], whereas the internal structure of a
Cluster consisted of particles distant $\sim$0.3 [nm].

In this way the work points quite unequivocally to clustering
processes and to a cluster-like structure of the matter and thus
contributes to the understanding of the structure of density
fluctuations and simultaneously sheds more light on the structure of
the universe in the moment when the universe became transparent for
photons. Clustering may be at the same time a new physical effect
which has not been taken fully into consideration in the past. On the
basis of our quantitative considerations it was possible to derive the
number of particles (protons, helium nuclei, electrons and other
particles) in Objects and Clusters and the number of Clusters in an
Object.

\begin{acknowledgements}
My thanks are due to Mgr. Radom\'{\i}r \v{S}m\'{\i}da, PhD (Institute
of Physics, Acad. Sci. of the Czech Republic) for comments, proposals
and discussion concerning this article, to Prof. Karel Segeth
(Institute of Mathematics, Acad. Sci. of the Czech Republic) for
discussions and help in clarifying some aspects of the Fourier
transform, to Prof. Jan Kratochv\'{\i}l (Department of Physics,
Faculty of Civil Engineering, Czech Technical University in Prague)
for discussions pointing out several inconsistencies in the original
conception of the article and to Prof. Richard Gerber (University of
Salford, Manchester) for discussion and proposals directed to the
final presentation of this paper. Last acknowledgement is due to
Dr. Ji\v{r}\'{\i} Hybler (Institute of Physics, Acad. Sci. of the
Czech Republic) for help in the preparation of Fig. 13 presenting a
part of a Cluster skeleton.
\end{acknowledgements}



\appendix

\section{Basic equations}

Generally the intensity of radiation scattered4 on a matter (solid,
liquid) offers us information on the structure of a material of any
kind in the reciprocal space. The relation between the reciprocal and
real space is mediated by the Fourier transform of the radiation
intensity scattered by a disordered material.

The basic formula transforming the reciprocal space information into
the real space one is in the case of non-crystalline (non-periodic)
materials (Steeb 1968)

\begin{equation}
\rho(r)  = 4 \pi r^{2} \sum_{m} a_{m} K_{m} \rho_{m}^{el}(r) \;\; .
\label{eqA1}
\end{equation}

In a more detailed description the quantity $\rho(r)$ is then expressed as

\begin{eqnarray}
\rho(r) & = & 4 \pi r^{2} \sum_{m} a_{m} K_{m} \rho_{0}^{el} 
\nonumber \\ 
 & & + \:
\frac{2r}{\pi} \int^{S^{\mathrm{max}}}_{0} \!\! s \, i(s) \, \sin(rs) \,
\exp(\tau s^{2}) \, ds
\label{eqA2}
\end{eqnarray}

and describes the radial electron density distribution function (RDF)
in real space in the case when the atomic scattering factor $f_{m}$
(see equation (A.6)) is given in electrons [e]. The parameter $r$ is
the distance of an arbitrary atom from the origin {\em in real space
  units}.

In equation (A.2) $a_{m}$\ are the concentrations of elements
composing the matter ($\sum_{m}a_{m}=1$), $\rho_{m}^{el}(r)$\ are the
elemental contributions of electron density to the overall electron
density, i.e. it is the electron density around an atom of kind $m$,
the factor $\exp(\tau s^{2})$\ is an artificial temperature factor in
which usually $\tau$=-0.010, $\rho_{0}^{el}$\ is the mean electron
density in a totally disordered material, which can be deduced from
the macroscopic density via the Avogadro number $L$

\begin{equation}
\rho_{0}^{el} = \frac{L}{M} D \times 10^{-21} \times
\sum_{m}a_{m}Z_{m}
\label{eqA3}
\end{equation}

where $Z_{m}$\ is the atomic number of kind m, $D$ is the
macroscopic density in [g.cm$^{-3}$] and $M$ is the molecular weight

\begin{equation}
M = \sum_{m} a_{m} W_{m}
\label{eqA4}
\end{equation}

$W_{m}$\ are corresponding atomic weights. The factor 10$^{-21}$\ in
equation (A.3) is a consequence of the fact that the parameter $r$\ is
in [nm].

The parameter $s$ is in equation (A.2) related with the wavelength of
scattered radiation $\lambda$ by the formula

\begin{equation}
s = 4 \pi \frac{\sin \theta}{\lambda}
\label{eqA5}
\end{equation}

Here is $s = \mathbf{s}-\mathbf{s_{0}}$, where $\mathbf{s_{0}}$\ is
the vector of the incident and $\mathbf{s}$\ the vector of the
scattered radiation in the reciprocal space.

Further, $\theta$\ is the angle between the incident and scattered
radiation (X-rays or neutrons) and $\lambda$ is the wavelength of this
radiation and $K_{m}$ is the effective number of electrons in an atom
of kind m

\begin{equation}
K_{m} = f_{m} / f_{e}
\label{eqA6}
\end{equation}

where $f_{m}$ is the atomic scattering factor for X-rays for an atom
of the kind m (see already Sect.\ 2.1.1.) and and $f_{e}$ is the
atomic scattering power of an electron for X-rays

\begin{equation}
f_{e} = \frac{\sum_{m} a_{m} f_{m} }{\sum_{m} a_{m} Z_{m}}
\label{eqA7}
\end{equation}

During a conventional experiment (e.g. see Fig. 2), i.e. using
MoK$_{\alpha}$\ radiation, we have
$\lambda^{\mathrm{Mo}}_{\mathrm{Classic}}$\ = 0.071069 [nm] and the
maximum possible value of $s^{\mathrm{max}}_{\mathrm{Classic}}$ ,
corresponding to $\theta = 90^{\circ}$ is then according equation
(A.5)
\begin{equation}
s^{\mathrm{max}}_{\mathrm{Classic}} =
\frac{4\pi}{\lambda^{\mathrm{Mo}}_{\mathrm{Classic}}} = 176.819 
 [\mathrm{nm}^{-1}]
\label{eqA8}
\end{equation}

Here we are starting to use the subscript ``Classic'', which should
point out that the scattering vector in the reciprocal space
$s_{\mathrm{Classic}}$ will be considered in the same way as in the
``classic'' conventional non-crystalline case.

In equation (A.2) is $i(s)$ the experimentally obtained scattered
intensity of radiation, $I_{\mathrm{corr}}$ is this intensity
corrected for various factors
\footnote{In a conventional experiment the scattered intensity is
  corrected for scattering on ``air'', absorption, divergency of the
  X-ray beam, Lorentz and polarization factor. During our calculations
  we have included only the polarization factor.}  and properly scaled
for the absolute value of scattering, hence
\begin{equation}
i(s) = I_{\mathrm{distr}} = \left( I_{\mathrm{corr}}(s) - I_{\mathrm{gas}}(s) \right) / f^{2}_{e}
\;\; ,
\label{eqA9}
\end{equation}

the parameter $f^{2}_{e}$ is acting here as a sharpening function.

The general formula for the scattering on gas $I_{\mathrm{gas}}(s)$ is

\begin{equation}
I_{\mathrm{gas}}(s) = \left( \sum_{m} a_{m} f^{2}_{m} +  \sum_{m} a_{m}
f_{m}^{\mathrm{incoh}} 
\right) \;\; ,
\label{eqA10}
\end{equation}

where $f_{m}^{\mathrm{incoh}}$ are the scattering factors for
the incoherent (Compton) scattering, see Fig. 7.

The labelling $I_{\mathrm{distr}}$ for $i(s)$ will be used in the Appendix B,
where the scaling methods, important for a correct Fourier transform,
are discussed.


\section{The scaling problem}

In equation (A.9) we have already introduced the quantity
$I_{\mathrm{corr}}(s)$, i.e. the corrected experimental scattered
intensity. However, in order to arrive to a correct RDF,
$I_{\mathrm{corr}}(s)$ must be scaled to the $I_{\mathrm{gas}}(s)$
function in the absolute scale of atomic scattering, see equation
(A.10).

In the simplest scaling method we suppose that for high ${s}$-values
(HSV) there are not any scattering effects on the corrected
experimental curve $I_{\mathrm{corr}}(s)$ and therefore the
$I_{\mathrm{corr}}(s)$ and the $I_{\mathrm{gas}}(s)$ curves should be
equal. Then the scaling parameter $a_{\mathrm{HSV}}$ is for $s
\rightarrow s_{\mathrm{max}}$ easily calculated as

\begin{equation}
a_{\mathrm{HSV}} = \frac{I_{\mathrm{gas}}(s)}{I_{\mathrm{corr}}(s)}
\label{eqB1}
\end{equation}

As a consequence we obtain in the whole interval of $s$-values a
scaled scattered intensity $I_{\mathrm{norm}}^{\mathrm{HSV}}(s)$
represented by the equation

\begin{equation}
I_{\mathrm{norm}}^{\mathrm{HSV}}(s) = a_{\mathrm{HSV}} \, 
I_{\mathrm{corr}}(s)
\label{eqB2}
\end{equation}

The function $I^{\mathrm{HSV}}_{\mathrm{norm}}(s)$ oscillates around
the $I_{\mathrm{gas}}(s)$ curve. Following equation (A.9), we subtract
the scattering on gas and obtain the most important function
$I_{\mathrm{distr}}$, see Fig. 6.

There are several other scaling methods. An integral method according
to Hultgren et al. (1935) is characterized by a scaling factor
$a_{\mathrm{HGW}}$ and supposes that the areas under the experimental
scattering curve $I_{\mathrm{corr}}(s)$ and the structureless
$I_{\mathrm{gas}}(s)$ curve should be equal. Similarly there is a
quadratic integral method according to Krog Moe (1956) characterized
by a scaling parameter $a_{\mathrm{KRM}}$.

Our long experience in the research of disordered materials documents
that the better was the experiment and the better has been the
application of scattering factors, the smaller was the difference
(only several percent) between the scaling coefficients
$a_{\mathrm{HSV}}$, $a_{\mathrm{HGW}}$ and $a_{\mathrm{KRM}}$ and the smaller
were the parasitic fluctuations on the RDF. In the present work we
have used all three scaling methods and have kept the difference
between scaling factors in the limit of 4 percent.


\begin{thebibliography}{}

\bibitem[1999]{brandenburg} Brandenburg, K. 1999,
	Program DIAMOND, Version 2.1c, Crystal Impact GbR,
	Bonn 1999, Germany

\bibitem[1998]{cervinka98} \v{C}ervinka, L. 1998,
	J. of Non-Crystalline Solids, 232-234, 1

\bibitem[2005]{cervinka05} \v{C}ervinka, L., Bergerov\'{a},
J., Tich\'{y}, L. \& Rocca, F. 2005,
	Phys. \& Chem. of Glasses, 46, 444

\bibitem[2003]{hinshaw} Hinshaw, G., Spergel, D. N., Verde, L.,
Hill, R. S., Meyer, S. S., Barnes, C., Bennett, C. L., Halpern, M.,
Jarosik, N., Kogut, A., Komatsu, E., Limon, M., Page, L., Tucker, G. S.,
Weiland, J., Wollack E. \& Wright, E. L. 2003,
	Astrophys. J. Suppl., 148, 135

\bibitem[1995]{hu} Hu, W., Scott, D., Sugiyama, N. \& White, M. 1995,
	Phys. Rev. D52, 5498

\bibitem[1935]{hultgren} Hultgren, R., Gingrich,
N. S. \& Warren, B. E. 1935,
	J. Chem. Phys. 3, 351.

\bibitem[1956]{kroghmoe} Krogh Moe, J. 1956,
	Acta Crystallogr. 9, 951.

\bibitem[1994]{mather} Mather, J. C., Cheng, E. S., Cottingham, D. A.,
Eplee, R. E. Jr., Fixen, D. J., Hewagama, T., Isaacman, R. B.,
Jensen, K. A., Meyer, S. S., Noerdlinger, P. D., Read, S. M., 
Rosen, L. P., Shafer, R. A., Wright, E. L., Bennett, C. L., Boggess, N. W., 
Hauser, M. G., Kelsall, T., Moseley, S. H.Jr., Silverberg, R.F .,
Smoot, G. F., Weiss, R. \& Wilkinson, D. T. 1994,
	Astrophys. J. 420, 439

\bibitem[2006]{petricek} Pet\v{r}\'{\i}\v{c}ek, V.,
Du\v{s}ek, M. \& Palatinus, L. 2006,
	Jana2006 - The crystallographic computing system,
	Institute of Physics, Praha 2006, Czech Republic

\bibitem[2003]{sievers} Sievers, J. L., Bond, J. R., Cartwright, J. K.,
Contaldi, C. R., Mason, B. S., Myers, S. T., Padin, S.,
Pearson, T. J., Pen,U. L., Pogosyan, D., Prunet, S., Readhead, A. C. S.,
Shepherd, M. C., Udomprasert, P. S., Bronfman, L., Holzapfel, W. L. \&
May, J. 2003,
	Astrophys. J., 591, 592

\bibitem[1977]{silk} Silk, J. 1977,
	in Big Bang,
	(Freeman \& Co. Publishers, New York), 299

\bibitem[2007]{smolin} Smolin, L. 2007,
	in The Trouble with Physics,
	(Mariner Books, ISBN 061891868X), 16

\bibitem[1977]{smoot} Smoot, G. \& Davidson, K. 1977,
	in Wrinkles in Time,
	(Avon, New York), 158

\bibitem[1968]{steeb} Steeb, K. 1968,
	Springer Tracts in Modern Physics - 
	Ergebnisse der exakten Naturwissenschaften, Vol. 47, Editor G. Hohler,
	(Springer-Verlag, Berlin - Heidelberg - New York), 1.

\bibitem[2010]{smida} \v{S}m\'{\i}da, R.,
	Institute of Physics, Acad. Sci. of the Czech Rep.,
	2010, private communication

\bibitem[2007]{wheeler} Wheeler, J.C. 2007,
in Cosmic Catastrophes,
(Cambridge University Press, ISBN 0521857147), 282

\bibitem[1999]{wilson} Wilson, J.C. \& Price, E., Editors 1999,
	in International Tables for Crystallography,
	Volume C, Mathematical, physical and chemical tables, Second edition,
	Published for International Union of Crystallography
	(Kluwer Academic Publishers, Dordrecht - Boston - London)

\bibitem[1994]{wright} Wright, E. L., Mather, J. C., Fixsen, D. J.,
Kogut, A., Shafer, R. A., Bennett, C. L., Boggess, N. W., Cheng, E. S.,
Silverberg, R .F., Smoot, G. F. \& Weiss, R. 1994,
	Astrophys. J. 420, 450
\end{thebibliography}
\end{document}